\documentclass{aa}
\usepackage{natbib}
\bibpunct{(}{)}{;}{a}{}{,} 

\usepackage{graphicx}
\usepackage[dvipsnames]{xcolor}
\usepackage{txfonts}
\usepackage{comment}
\usepackage{array}
\usepackage{multirow}
\usepackage{hyperref}
\hypersetup{
    colorlinks=true,
    citecolor=blue,
    linkcolor=blue,
    urlcolor=blue
    }

\usepackage{tabularx}
\usepackage{newtxtext,newtxmath,amsmath}
\usepackage{makecell}
\usepackage{placeins}
\usepackage{lineno}

\renewcommand\makeLineNumber{}

\begin{document} 

   \title{Impact of \textit{T}- and $\rho$-dependent decay rates and new (n,$\gamma$) cross sections on the \textit{s} process in low-mass AGB stars}

   \author{B. Sz\'anyi
          \inst{1,2,3}
          \and
          A. Yag\"ue L\'opez\inst{4}
          \and
          A. I. Karakas\inst{5,6}
          \and
          M. Lugaro\inst{2,3,5,7}
          }

   \institute{Department of Experimental Physics, Institute of Physics, University of Szeged, H-6720 Szeged, Dóm tér 9, Hungary
         \and
             Konkoly Observatory, HUN-REN CSFK/RCAES, Konkoly Thege Mikl\'os \'ut 15-17, H-1121 Budapest, Hungary \\
             \email{szanyi.balazs@csfk.org}
             \and
                CSFK, MTA Centre of Excellence, Budapest, Konkoly Thege Mikl\'os út 15-17., H-1121, Hungary
          \and 
          Computer, Computational and Statistical Sciences (CCS) Division, Center for Theoretical Astrophysics, Los Alamos National Laboratory, Los Alamos, NM, 87545, USA 
          \and
          School of Physics and Astronomy, Monash University, VIC 3800, Australia
          \and
          ARC Centre of Excellence for All Sky Astrophysics in 3 Dimensions (ASTRO 3D)
          \and
          ELTE E\"{o}tv\"{o}s Lor\'and University, Institute of Physics and Astronomy, Budapest 1117, P\'azm\'any P\'eter s\'et\'any 1/A, Hungary
             }

   \date{Received 20 November 2024; accepted 1 March 2025}

 
  \abstract
   {}
   {We study the impact of nuclear input related to weak-decay rates and neutron-capture reactions on predictions for the \textit{slow} neutron-capture process (\textit{s} process) in asymptotic giant branch (AGB) stars. We provide the first database of surface abundances and stellar yields of the isotopes heavier than iron from the \textit{Monash} models.}
   {We run nucleosynthesis calculations with the \textit{Monash} post-processing code for seven stellar structure evolution models of low-mass AGB stars \iffalse(2, 3 and 4 $\mathrm{M_{\odot}}$ with solar metallicity and 3 and 4 $\mathrm{M_{\odot}}$ with half-solar and double-solar metallicity)\fi with three different sets of nuclear input. The reference set has constant decay rates and represents the set used in the previous \textit{Monash} publications. The second set contains the temperature and density dependence of $\mathrm{\beta}$ decays and electron captures based on the default rates of nuclear NETwork GENerator (NETGEN). In the third set, we further update 92 neutron-capture rates based on reevaluated experimental cross sections from the ASTrophysical Rate and rAw data Library. We compare and discuss the predictions of each set relative to each other in terms of isotopic surface abundances and total stellar yields. We also compare results to isotopic ratios measured in presolar stardust silicon carbide (SiC) grains from AGB stars.}
   {The new sets of models resulted in a $\sim$66\% solar \textit{s}-process contribution to the \textit{p}-nucleus $\mathrm{^{152}Gd}$, confirming that this isotope is predominantly made by the \textit{s} process. The nuclear input updates resulted in predictions for the $\mathrm{^{80}Kr}$/$\mathrm{^{82}Kr}$ ratio in the He intershell and surface $\mathrm{^{64}Ni}$/$\mathrm{^{58}Ni}$, $\mathrm{^{94}Mo}$/$\mathrm{^{96}Mo}$ and $\mathrm{^{137}Ba}$/$\mathrm{^{136}Ba}$ ratios more consistent with the corresponding ratios measured in stardust, however, the new predicted $\mathrm{^{138}Ba}$/$\mathrm{^{136}Ba}$ ratios are higher than the typical values of the stardust SiC grain data. The W isotopic anomalies are in agreement with data from analysis of other meteoritic inclusions. We confirm that the production of $\mathrm{^{176}Lu}$ and $\mathrm{^{205}Pb}$ is affected by too large uncertainties in their decay rates from NETGEN.}
   {}


   \maketitle
%

\section{Introduction}

At the end of their lives, low-mass stars (1 $\mathrm{M_{\odot}}$ $\lesssim$ \textit{M} $\lesssim$ 8 $\mathrm{M_{\odot}}$) evolve to the asymptotic giant branch \citep[AGB,][]{ib84,bu99,he04,la14}. AGB stars are characterised by a degenerate C-O core, a H and a He shell, and an extensive, convective, H-rich envelope. The two shells are separated by a thin He-rich layer called the intershell. The energy production occurs in cycles: a relatively long phase of H burning is periodically (every $\sim$ $10^3$-$10^5$ years, depending on the stellar mass) interrupted by a thermal instability of the He shell, which is called a thermal pulse \citep[TP,][]{sc65, we66}. The rapid release of a substantial amount of energy during a TP causes the development of a convective region in the intershell, hereafter referred to as a "convective pulse", and the star to expand from the He shell outward. This expansion allows the envelope to extend downward after the extinction of the convective pulse, potentially penetrating the intershell region and carrying material from there to the surface. The resulting mixing episodes are collectively referred to as the third dredge-up (TDU). The intershell material carried to the stellar surface contains a large amount of carbon produced by partial He burning, therefore the TDU can convert the surface layers of the AGB star to evolve from oxygen rich to carbon rich \citep[C/O$\ge$1,][]{wa98}. The cycle described above is repeated several times (tens to hundreds of times, depending on the initial mass) before the envelope is completely eroded by mass loss, and the C-O core of the star is left as a white dwarf.

In the thermally pulsing AGB phase, conditions are suitable for ($\mathrm{\alpha}$,n) neutron-source reactions to be initiated in the intershell. The $\mathrm{^{22}Ne(\alpha,n)^{25}Mg}$ reaction \citep[][]{ca60,tr77} is activated for temperatures greater than 300 MK during the convective pulses of AGB stars with an initial mass greater than $\sim$ 3 $\mathrm{M_{\odot}}$ and results in a high maximum neutron density \citep[$\sim$ $10^{13}$ $\mathrm{n\,cm^{-3}}$, e.g.][]{fi14} but a low total amount of free neutrons. The $\mathrm{^{13}C(\alpha,n)^{16}O}$ reaction \citep[initially reported by][]{ho88} is activated during the H-burning phase, at \textit{T} $\sim$ 90 MK and results in a high total amount of free neutrons, but a low maximum neutron density \citep[$\sim$ $\mathrm{10^{7}}$ $\mathrm{n\,cm^{-3}}$,][]{st95}. The captures of the resulting free neutrons drive \textit{slow} neutron captures, that is, the \textit{s} process \citep{bu57,ka11,lu23}, which produces roughly half of the abundances of elements heavier than iron (up to lead) in the Universe \citep[see review by][]{lu23}.

During the \textit{s} process, by definition, the probability of neutron capture on an unstable isotope is lower than its decay, therefore, the main neutron-capture path follows the valley of $\mathrm{\beta}$-stability. However, depending on the temperature, density, and neutron density, the neutron-capture rate of some unstable isotope may compete with their decay rate. These unstable isotopes are known as branching points on the \textit{s}-process path \citep[e.g.][]{bi15,lu18}. The branching factor ${f_{\mathrm{\beta}}}$ is defined as:
\begin{equation}
\label{eq:1}
f_{\mathrm{\beta}} = 1 - f_{\mathrm{n}} = \mathrm{\frac{\lambda_{\beta}}{\lambda_{\beta} + \lambda_n}},
\end{equation}
where $f_{\mathrm{n}}$ is the branching factor of the neutron capture channel, $\mathrm{\lambda_{\beta}}$ is the $\beta$-decay rate ($\mathrm{\lambda_{\beta}}$ = ln2\,$t_{1/2}^{-1}$, where $t_{1/2}$ is the half-life) and $\mathrm{\lambda_n}$ = $N_\mathrm{n}\,\langle \sigma v \rangle$ is the neutron-capture rate given by the neutron density $N_\mathrm{n}$, the relative velocity of neutrons and target nuclei $v$, and the neutron capture cross section $\sigma$. The $\langle \sigma v \rangle$ can be approximated as the product of the Maxwellian-average cross section (MACS or $\sigma_\mathrm{M}$) and the thermal velocity $v_{\mathrm{th}}$. 

The calculation of $f_{\mathrm{\beta}}$ is not trivial because not only $N_\mathrm{n}$ varies, but also $\mathrm{\lambda_{\beta}}$ and $\sigma_\mathrm{M}$ can change with \textit{T} and $\rho$. Neutron-capture rates on a nucleus ground state do not significantly vary with temperature, however, to obtain neutron-capture rates in a stellar environment, in addition to laboratory ground-state cross sections, the theoretical relative contribution of excited states \citep{ra12}, generally described using a stellar enhancement factor (SEF), needs to be included. Also, laboratory $\mathrm{\beta}$-decay rates can be largely modified at stellar temperatures also due to the thermal population of low-lying excited states, which may have different $\beta$-decay lifetimes \citep[e.g.][]{re18b}. In addition, temperature and electron density change the population of electronic states, providing the possibility of the bound-state decay \citep{ba61}. Electron-capture rates are also influenced by temperature and electron density: the likelihood of a bound electron near the nucleus decreases with temperature, while the probability of a free electron near the nucleus increases with electron density \citep[e.g.][]{ta87}. Accounting for these effects is essential for the accurate modelling of the \textit{s} process in AGB stars. 

Here we present a major upgrade of nuclear reaction rates of the \textit{Monash} nucleosynthesis code. Most of the radioactive-decay and electron-capture reactions were constant which we replace with \textit{T}- and $\rho$-dependent tables. These tables were based on the compilation of NETGEN \citep[nuclear NETwork GENerator,][]{jo01,ai05,xu13}. Most other sets of models -- NuGrid \citep[Nucleosynthesis Grid, e.g.][]{pi12}; FUNS \citep[FUll Network Stellar, e.g.][]{st06} and STAREVOL \citep[e.g.][]{si06} -- already have implemented \textit{T}- and $\rho$-dependent decay rates and provided the resulting yields and abundances. However, the thorough analysis that we present here regarding the impact of using variable weak rates on the activation of branching points and the implications on the comparison with observations, including the interpretation of stardust grain data is still missing in the literature. To our knowledge, recently only \citet{bi15} presented a detailed analysis of branching points, but from the point of view of the impact of the uncertainties of the neutron source reactions. Furthermore, we include new tabulated neutron-capture rates for 92 reactions using reevaluated experimental MACSs from the ASTRAL database \citep[ASTrophysical Rate and rAw data Library,][]{re18,ve21,ve23} and SEFs from KaDoNiS \citep[Karlsruhe Astrophysical Database of Nucleosynthesis in Stars,][]{di06}. We provide the first database of surface abundances and stellar yields of the isotopes heavier than iron from the \textit{Monash} models.

\setlength{\tabcolsep}{1.0em}
\begin{table}
\caption{Relevant features of the stellar models grouped by metallicity.}
\label{table:table 1}
\centering
\begin{tabular}{ccccc}
\hline 
\hline
\makecell{\textit{M} \\ ($\mathrm{M_{\odot}}$)} & \#TDU & \#$\mathrm{TP_{C/O>1}}$ & \makecell{$\mathrm{\textit{T}^{max}_{TP}}$ \\ (MK)} & \makecell{$M_{\mathrm{PMZ}}$ \\ ($\mathrm{M_{\odot}}$)} \\
\hline
\multicolumn{5}{c}{\textit{Z} = 0.007, \textit{Y} = 0.26} \\ 
\hline
 3.00 & 19 & 17 & 310 & 2 $\times$ $10^{-3}$ \\
 4.00 & 23 & 12 & 360 & 1 $\times$ $10^{-3}$ \\
\hline 
\multicolumn{5}{c}{\textit{Z} = 0.014, \textit{Y} = 0.28} \\ 
\hline
 2.00 & 9 & 2 & 280 & 2 $\times$ $10^{-3}$ \\
 3.00 & 17 & 10 & 302 & 2 $\times$ $10^{-3}$ \\
 4.00 & 20 & 8 & 348 & 1 $\times$ $10^{-3}$ \\
\hline 
\multicolumn{5}{c}{\textit{Z} = 0.03, \textit{Y} = 0.30} \\ 
\hline 
 3.00 & 17 & 2 & 294 & 2 $\times$ $10^{-3}$ \\
 4.00 & 21 & 1 & 324 & 1 $\times$ $10^{-3}$ \\
\hline
\end{tabular}
\tablefoot{Columns indicate the initial mass (\textit{M}), the number of third dredge-ups (\#TDU), the number of thermal pulses after the envelope became carbon rich (\#$\mathrm{TP_{C/O>1}}$), the maximum temperature reached during thermal pulses ($\mathrm{\textit{T}^{max}_{TP}}$), and the mass extent of the partial mixing zone ($M_{\mathrm{PMZ}}$).}
\end{table}

The paper is organised as follows. We start with a description of our methods and of the seven stellar evolutionary models used for the post-processing nucleosynthesis calculations (Sect. \ref{section:section 2}). In Sect. \ref{section:section 3}, we present the impact of nuclear input changes on nucleosynthesis predictions and discuss the results including a comparison to other studies and a selection of isotopic ratios measured in presolar stardust silicon carbide (SiC) grains from AGB stars. The conclusions are provided in Sect. \ref{section:section 4}.

\section{Methods and models}
\label{section:section 2}

We provide predictions for stellar abundances from AGB stars, using the \textit{Monash} post-processing nucleosynthesis code \citep{ca93}, which calculates the changes in the abundances of currently 328 nuclear species due to convective mixing and nuclear burning on the basis of detailed stellar structure calculated by the \textit{Monash} evolution code \citep{la86,ka14}. In summary, the current nuclear reaction network of the nucleosynthesis code includes 2\,351 reactions, most of which are based on the JINA reaclib database \citep{cy10}, where reaction rates are expressed as a function of temperature $T_9$ (in units of $10^9$ Kelvin) using a seven-parameter ($a_{0-6}$) analytical formula \citep{th86}: 
\begin{equation}
\lambda = \mathrm{exp}\left [a_0 + \sum_{\mathrm{i}=1}^{5} a_{\mathrm{i}}\,T_9^{\frac{2\mathrm{i}-5}{3}} + a_6\,\mathrm{ln}T_9\right ].
\end{equation}
While the JINA reaclib database is advantageous for building a large network, it has two major disadvantages: (1) For some experimental rates, there can be a variation of up to 10\%-20\% of the fit from the actual rate. (2) The $\mathrm{\beta}$-decay and electron-capture rates are constant, that is, do not depend on temperature and electron density. For these reasons, a new version of the nucleosynthesis code was developed by \citet{ra12}, which includes a routine that allows us to use specific reaction rates as tabulated values instead of the reaclib fit, and here it was extended to decay rates. 

\setlength{\tabcolsep}{0.64em}
\begin{table*}
\caption{The sources of the reaction rates in the different sets.}
\label{table:2}
\centering
\begin{tabular}{c c c}
\hline 
\hline
& $\mathrm{\beta}$-decay rates & neutron-capture rates \\
\hline
 & Constant/terrestrial rates from JINA reaclib\tablefootmark{a} & \textit{ka02} from JINA reaclib \tablefootmark{a}, except for \\
Set 0 & $\mathrm{^{134}Cs(\beta^-)^{134}Ba}$ from (1) & 24\tablefootmark{b} isotopes taken directly from KaDoNiS\tablefootmark{c} v0.2 (MACSs + SEFs), and \\
 & $\mathrm{^{181}Hf(\beta^-)^{181}Ta}$ from (2) & $\mathrm{^{90,91,92,93,94,95,96}Zr(n,\gamma)}$ from (3) \\
\hline
 & \textbf{NETGEN \tablefootmark{d}} for 102\tablefootmark{e} isotopes
 & \\
Set 1 & $\mathrm{^{134}Cs(\beta^-)^{134}Ba}$ from (1) & Same as above \\
 & $\mathrm{^{181}Hf(\beta^-)^{181}Ta}$ from (2) & \\
\hline
 & & \textit{ka02} from JINA reaclib \tablefootmark{a}, except for \\
Set 2 & Same as above & 18\tablefootmark{f} isotopes taken directly from KaDoNiS\tablefootmark{c} v0.2 (MACS + SEF), and \\
 & & 92\tablefootmark{g} from \textbf{ASTRAL v0.2}\tablefootmark{h}(MACS) and KaDoNiS\tablefootmark{c} v0.3 (SEF) \\
\hline
\end{tabular}
\tablebib{(1) \citet{li21}; (2) \citet{lu14a}; (3) \citet{lu14}}
\tablefoot{\\ \tablefoottext{a}{https://reaclib.jinaweb.org} \\ 
\tablefoottext{b}{for the 24 isotopes: $\mathrm{^{28}Si}$, $\mathrm{^{32}S}$, $\mathrm{^{40}Ca}$, $\mathrm{^{54}Cr}$, $\mathrm{^{58}Fe}$, $\mathrm{^{60,62,64}Ni}$, $\mathrm{^{138}Ba}$, $\mathrm{^{151,153}Eu}$, $\mathrm{^{158}Gd}$, $\mathrm{^{159}Tb}$, $\mathrm{^{182}Ta}$, $\mathrm{^{182,183,184,185,186}W}$, $\mathrm{^{186,187,188,189,190}Os}$} \\
\tablefoottext{c}{https://kadonis.org} \\ 
\tablefoottext{d}{http://www.astro.ulb.ac.be/Netgen/} \\ 
\tablefoottext{e}{\citet{ho96} for 35 isotopes: $\mathrm{^{14}C}$, $\mathrm{^{13}N}$, $\mathrm{^{14,15,19}O}$, $\mathrm{^{17,18,20}F}$, $\mathrm{^{19,23}Ne}$, $\mathrm{^{21,22,24}Na}$, $\mathrm{^{23,27}Mg}$, $\mathrm{^{25,26,26m,28}Al}$, $\mathrm{^{27}Si}$, $\mathrm{^{32,33}P}$, $\mathrm{^{35}S}$, $\mathrm{^{36}Cl(\beta^-)}$, $\mathrm{^{37,39}Ar}$, $\mathrm{^{40}K}$, $\mathrm{^{41,45,47}Ca}$, $\mathrm{^{46,47,48}Sc}$, $\mathrm{^{55}Fe}$, $\mathrm{^{85m}Kr}$ \\
\citet{cf88} for $\mathrm{^{7}Be}$, \citet{od94} for $\mathrm{^{36}Cl(\epsilon)}$ and \citet{la00} for $\mathrm{^{48}Ca}$ \\
\citet{go99} for 47 isotopes: $\mathrm{^{59,60}Fe}$, $\mathrm{^{60}Co}$, $\mathrm{^{59,63}Ni}$, $\mathrm{^{64}Cu}$, $\mathrm{^{65}Zn}$, $\mathrm{^{71}Ge}$, $\mathrm{^{79}Se}$, $\mathrm{^{81}Kr}$, $\mathrm{^{86}Rb(\beta^-)}$, $\mathrm{^{87}Rb}$, $\mathrm{^{93}Zr}$, $\mathrm{^{94}Nb}$, $\mathrm{^{99}Tc}$, $\mathrm{^{107}Pd}$, $\mathrm{^{108}Ag(\epsilon)}$, $\mathrm{^{109}Cd}$, $\mathrm{^{128,129}I}$, $\mathrm{^{134,135,137}Cs}$, $\mathrm{^{147}Nd}$, $\mathrm{^{147,148}Pm}$, $\mathrm{^{151}Sm}$, $\mathrm{^{152,154,155}Eu}$, $\mathrm{^{153}Gd}$, $\mathrm{^{161}Tb}$, $\mathrm{^{170}Tm(\beta^-)}$, $\mathrm{^{171}Tm}$, $\mathrm{^{177}Lu}$, $\mathrm{^{182}Hf}$, $\mathrm{^{182}Ta}$, $\mathrm{^{185}W}$, $\mathrm{^{186,187}Re}$, $\mathrm{^{191}Os}$, $\mathrm{^{192}Ir(\beta^+)}$, $\mathrm{^{193}Pt}$, $\mathrm{^{204,205}Tl}$, $\mathrm{^{205}Pb}$, $\mathrm{^{210}Bi}$} \\ \citet{ta87} for 22 isotopes: $\mathrm{^{69}Zn}$, $\mathrm{^{70}Ga}$, $\mathrm{^{80}Br}$, $\mathrm{^{85}Kr}$, $\mathrm{^{86}Rb(\epsilon)}$, $\mathrm{^{89,90}Sr}$, $\mathrm{^{90,91}Y}$, $\mathrm{^{95}Zr}$, $\mathrm{^{95}Nb}$, $\mathrm{^{108}Ag(\beta^-)}$, $\mathrm{^{133}Xe}$, $\mathrm{^{136}Cs}$, $\mathrm{^{141}Ce}$, $\mathrm{^{159}Gd}$, $\mathrm{^{161}Tb}$, $\mathrm{^{169}Er}$, $\mathrm{^{170}Tm(\epsilon)}$, $\mathrm{^{177}Lu}$, $\mathrm{^{183}Ta}$, $\mathrm{^{192}Ir(\beta^-)}$ \\
\tablefoottext{f}{As in (\textit{b}) minus $\mathrm{^{40}Ca}$, $\mathrm{^{58}Fe}$, $\mathrm{^{64}Ni}$, $\mathrm{^{158}Gd}$, $\mathrm{^{186}W}$, and $\mathrm{^{190}Os}$} \\
\tablefoottext{g}{for the 92 isotopes: $\mathrm{^{13}C}$, $\mathrm{^{23}Na}$, $\mathrm{^{36,38}Ar}$, $\mathrm{^{40}Ca}$, $\mathrm{^{45}Sc}$, $\mathrm{^{58}Fe}$, $\mathrm{^{59}Co}$, $\mathrm{^{64}Ni}$, $\mathrm{^{65}Cu}$, $\mathrm{^{64,70}Zn}$, $\mathrm{^{74}Ge}$, $\mathrm{^{75}As}$, $\mathrm{^{78}Se}$, $\mathrm{^{79,81}Br}$, $\mathrm{^{85}Rb}$, $\mathrm{^{89}Y}$, $\mathrm{^{103}Rh}$, $\mathrm{^{110,111,112,113,114}Cd}$, $\mathrm{^{116,117,118,120}Sn}$, $\mathrm{^{121}Sb}$, $\mathrm{^{122,123,124,125,126}Te}$, $\mathrm{^{128,129,130,132,134}Xe}$,$\mathrm{^{135}Cs}$, $\mathrm{^{134,135,136,137}Ba}$, $\mathrm{^{140,142}Ce}$, $\mathrm{^{141}Pr}$, $\mathrm{^{142,143,144,145,146,148}Nd}$, $\mathrm{^{148,149,150,151,152}Sm}$, $\mathrm{^{152,154,155,156,157,158}Gd}$, $\mathrm{^{160,161,162,163,164}Dy}$, $\mathrm{^{170}Er}$, $\mathrm{^{170,171,172,173,174}Yb}$, $\mathrm{^{175,176}Lu}$, $\mathrm{^{176,177,178,179,180,182}Hf}$, $\mathrm{^{181}Ta}$, $\mathrm{^{186}W}$, $\mathrm{^{187}Re}$, $\mathrm{^{190,192}Os}$, $\mathrm{^{191,193}Ir}$, $\mathrm{^{197}Au}$} \\
\tablefoottext{h}{https://exp-astro.de/astral/}}\\ 
\end{table*}

\subsection{Stellar models}
\label{section:section 2.1}

As input for post-processing nucleosynthesis calculations, we use a selection from the stellar structure evolution models presented in \citet{ka14} and \citet{ka16}. We select \textit{M} = 2, 3, 4 $\mathrm{M_{\odot}}$, \textit{Y} = 0.28 stellar models of solar metallicity \citep[\textit{Z} = 0.014, based on the solar abundances from][]{as09}, \textit{M} = 3, 4 $\mathrm{M_{\odot}}$, \textit{Y} = 0.26 stellar models at half-solar metallicity (\textit{Z} = 0.007), and \textit{M} = 3, 4 $\mathrm{M_{\odot}}$, \textit{Y} = 0.30 stellar models at double-solar metallicity (\textit{Z} = 0.03)\footnote{\textit{Y} indicates the mass fraction of helium and \textit{Z} indicates the mass fraction of metals, i. e. the metallicity of the stellar model.}. Input physics and methodology of the stellar structure models are described in full in the papers mentioned above. In summary, mass loss was not included in the red giant branch (RGB) phase and the mass-loss rate of \citet{va93} was used in the AGB phase. The models used the mixing length theory of convection with a mixing-length parameter $\mathrm{\alpha}$ = 1.86. The convective overshoot was not included in the calculations except for the 3 $\mathrm{M_{\odot}}$ \textit{Z} = 0.03 model, where an overshoot parameter ($N_{\mathrm{ov}}$ = 1) was used which resulted in a C-rich star.

Table \ref{table:table 1} shows some relevant properties of the calculated AGB models. (1) The total number of third dredge-up episodes (\#TDU) increases with increasing initial stellar mass. (2) The number of thermal pulses during which the envelope is C-rich depends also on the mass, the mass loss, and the activation of hot bottom burning (i. e. H-burning at the base of the envelope) in the 4 $\mathrm{M_{\odot}}$ stars. (3) The maximum temperature reached during thermal pulses, which controls the activation of the $\mathrm{^{22}Ne(\alpha,n)^{25}Mg}$ neutron source ($\mathrm{\textit{T}^{max}_{TP}}$) increases with increasing stellar mass and decreasing metallicity.

\subsection{Post-processing calculations}
\label{section:section 2.2}

The operation of the $\mathrm{^{13}C(\alpha,n)^{16}O}$ neutron source requires the formation of a partial mixing zone (PMZ), where protons from the convective envelope are mixed with the intershell matter and captured by the abundant $\mathrm{^{12}C}$, resulting in the so-called $\mathrm{^{13}C}$ pocket \citep{st97, ga98, go00}. Our method for including the $\mathrm{^{13}C}$ pocket is described in detail in \citet{ka16}. In summary, protons are partially mixed at the end of each TDU down to a mass extent in the intershell denoted by $M_{\mathrm{PMZ}}$, using an exponential profile. At the top of the intershell (corresponding to the base of the envelope during the TDU), the abundance of protons $\mathrm{\textit{X}_p}$ is that of the envelope, while at $M_{\mathrm{PMZ}}$ in the intershell it is $\mathrm{\textit{X}_p}$ = $10^{-4}$. Below this mass point, $\mathrm{\textit{X}_p}$ = 0. Different $M_{\mathrm{PMZ}}$ values as functions of the stellar mass are defined and justified by \citet{ka16} and \citet{bu17}. Here we use the "standard" values as reported in Table \ref{table:table 1}. 

We run post-processing nucleosynthesis calculations with three different inputs (see Table \ref{table:2}) for the nuclear network. The original set (hereafter Set 0) contains nuclear reactions mostly from the JINA reaclib database. The neutron-source reactions
$\mathrm{^{13}C(\alpha,n)^{16}O}$ and $\mathrm{^{22}Ne(\alpha,n)^{25}Mg}$ are from \citet{he08} and \citet{il10}, respectively. For neutron capture reactions, Set 0 uses the \textit{ka02} labelled fits of JINA reaclib, which are based on the KaDoNiS database, except for Zr isotopes, whose rates are from \citet{lu14}, and for a few other isotopes, whose rates are taken directly from KaDoNiS. These were selected because the JINA fit did not accurately cover the KaDoNiS data. This is the same network used by \citet{ka16}, except that in Set 0 we updated the temperature dependence of the $\mathrm{\beta}$-decay rate of $\mathrm{^{134}Cs}$ from \citet{ta87} used in \citet{ka16} to \citet{li21} and of $\mathrm{^{181}Hf}$ from \citet{lu14a}, which was included as terrestrial constant in \citet{ka16}.

In the second set (hereafter Set 1), we implement the temperature and density dependence of $\mathrm{\beta}$-decays and electron captures in the network using the default (most recent) rates from the NETGEN library. Below mass number \textit{A} = 56, the rates are based on \citet{ho96}, except for the $\mathrm{\beta^-}$-decay rate of $\mathrm{^{48}Ca}$, which is based on \citet{la00}, and the electron-capture rates of $\mathrm{^{7}Be}$ and $\mathrm{^{36}Cl}$, which are based on \citet{cf88} and \citet{od94}, respectively. Above \textit{A} = 56, the rates are based on the nominal rates of \citet{go99}, while the reactions not included in \citet{go99} are based on \citet{ta87}, except for the $\mathrm{\beta^-}$-decay rate of the excited $\mathrm{^{85}Kr}$ isomer, which is based on \citet{ho96}. In addition, we extend the reaction list by including the bound-state $\mathrm{\beta}$ decay of the terrestrial stable $\mathrm{^{205}Tl}$ based on \citet{go99}. Most of our decay tables contain the values in a grid of 250 temperature points between 1 MK and 1 GK on a logarithmic scale. For the rates that also depend on the electron density, the NETGEN tables consist of data in the same temperature grid as above for four electron densities: 1, 3, 9.5 or 10 (depending on the reference), and 30 $\times$ $\mathrm{10^{26}}$ $\mathrm{cm^{-3}}$, except for $\mathrm{^{36}Cl}$ and $\mathrm{^{48}Ca}$, which include data for 11 electron density points between 6 $\times$ $\mathrm{10^{24}}$ and 6 $\times$ $\mathrm{10^{33}}$ $\mathrm{cm^{-3}}$. We tested the possible effect of the choice of temperature grid on the \textit{M} = 3 $\mathrm{M_{\odot}}$, \textit{Z} = 0.014 model. We found that the application of a looser grid (125 points) or a finer grid (300 points) results in the same abundances within 1\%, except for $\mathrm{^{204}Pb}$, $\mathrm{^{205}Pb}$, and $\mathrm{^{205}Tl}$, which changed by less than 10\%. 

In the third set (hereafter Set 2), we upgrade the neutron-capture rates of 92 reactions using re-evaluated experimental MACSs from the ASTRAL database and SEFs from KaDoNiS v0.3. The current version (v0.2) of ASTRAL includes MACSs for 122 isotopes based on time-of-flight measurements, which have been re-evaluated by taking into account the new recommended value for the gold differential cross section \citep{re18} and 50 isotopes whose experimental cross section data are based on activation measurements \citep{ve23}. The neutron-capture rates at each temperature are calculated as $\lambda_n$ = $\left ( N_A\,\sigma_{\mathrm{M}}\,v_{\mathrm{th}} \right ) \times \mathrm{SEF}$, where $N_A$ = 6.02 $\times$ $10^{23}$ is the Avogadro constant. 

\subsection{Surface abundances, stellar yields and \texorpdfstring{$\delta$}{} values}
\label{section:section 2.3}

The nucleosynthesis code provides us with the time-evolution of normalised number abundances $Y_{\mathrm{k}}^{\mathrm{i}}$ of each isotope \textit{i} at the surface of the model stars. Using these outputs, we can calculate the total stellar yield $M_{\mathrm{i}}$ as the total mass of each isotope \textit{i} expressed in unit of solar mass, ejected into the interstellar medium (ISM) during the entire life of the star. To do this, we use an approximate formula, which sums over all the thermal pulses (instead integrating over the entire lifetime of the star) the mass lost for each isotope in-between each thermal pulse (identified by the index \textit{k}) according to: 
\begin{equation}
\label{eq:4}
M_{\mathrm{i}} = \left [ \mathrm{\sum^{final}_{k=1}} \left ( M_{\mathrm{k-1}} - M_{\mathrm{k}} \right ) X^{\mathrm{i}}_{\mathrm{k}} \right ] + \left ( M_{\mathrm{final}} - M_{\mathrm{core}} \right ) X^{\mathrm{i}}_{\mathrm{final}},
\end{equation}
where $M_\mathrm{k}$ is the mass of the model star at the end of thermal pulse \textit{k}, $X_{\mathrm{k}}^{\mathrm{i}}$ = $Y_{\mathrm{k}}^{\mathrm{i}}m_{\mathrm{i}}$ is the surface abundance of isotope $i$ in mass fraction at the end of same thermal pulse \textit{k} (where $m_{\mathrm{i}}$ is the atomic mass of isotope \textit{i}), $M_{\mathrm{final}}$ and $M_{\mathrm{core}}$ are the total mass and the core mass at the end of the computed evolution (i. e. where the code stops converging). The second term of the equation is added to remove the full envelope, assuming that no more TDU happens. We provide $Y_{\mathrm{k}}^{\mathrm{i}}$ and $M_{\mathrm{i}}$ for all our models as \href{https://zenodo.org/records/14981333}{Online Material}. Note that this approximated formula works here because the models we consider do not have any mass loss on the RGB and we tested that our Set 0 yields are within 2\% of those published by \citet{ka16}.

While stellar yields are an important input for the Galactic chemical evolution (GCE) calculations \citep[e.g.][]{co19,tr22}, isotopic abundance ratios are needed to interpret the composition of presolar stardust grains, such as SiC grains found in meteorites. Most of the SiC grains condensed in carbon-rich AGB stellar winds and were ejected into the ISM \citep{zi14}. The grains were transported to the protosolar nebula, trapped within the parent bodies of meteorites, which delivered them to Earth where they can be extracted and analysed in the laboratory \citep[e.g.][]{be87,le90}. The Presolar Grain Database \citep[PGD,][]{hy09,st24} contains all the available isotope data for single presolar stardust SiC grains. To compare to the SiC grain data, we calculate isotopic ratios of each element in the form of $\mathrm{\delta}$ values, that is, per mil variations with respect to the solar ratio as:
\begin{equation}
\mathrm{\delta_{k}{^iE}_j} = \left (\frac{\left( Y_{\mathrm{k}}^{\mathrm{i}}/Y_{\mathrm{k}}^{\mathrm{j}} \right )}{\left (Y_{\mathrm{k}}^{\mathrm{i}}/Y_{\mathrm{k}}^{\mathrm{i}} \right )_{\mathrm{\odot}} } -1 \right ) \mathrm{\times 1000},
\end{equation}
where $i$,$j$ represents the isotopes of element $E$ at the end of thermal pulse \textit{k}. The solar abundances in this case are also based on \citet{as09}.

\section{Results and discussion}
\label{section:section 3}
\subsection{Comparison of Set 1 to Set 0}
\label{section:section 3.1.}

In this section, we compare the predictions of the Set 0 and Set 1 models using the final surface normalised number abundances $Y^{\mathrm{i}}_{\mathrm{final}}$ of the model stars. There are 30 isotopes whose abundance changed by at least 10\% in at least one Set 1 stellar model compared to the corresponding Set 0 stellar model. The list of these isotopes and their $Y^\mathrm{i}_{\mathrm{Set\,1,final}}$/$Y^\mathrm{i}_{\mathrm{Set\,0,final}}$ ratio are shown in Table \ref{table:A.1.}. The abundances of individual isotopes, in addition to the nuclear input, depend on the characteristics of the stellar models (e.g. temperature, electron and neutron density, number of mixing episodes) and therefore the effects of the new rates can differ greatly between models. In the following, we present the most relevant isotope cases. 

Table \ref{table:table 3} highlights the relevant information for each case, including the branching isotopes, the decay modes and decay rates of the corresponding isotopes, the direction in which the abundance of each isotope has changed in the Set 1 models compared to Set 0 models, and the relevance of each case.

\subsubsection{\texorpdfstring{$\mathrm{^{64}Ni}$}{}}
\label{section:section 3.1.1.}

The \textit{s}-process production of $\mathrm{^{64}Ni}$ depends on the operation of the $\mathrm{^{63}Ni}$ and $\mathrm{^{64}Cu}$ branching points. $\mathrm{^{63}Ni}$ suffers $\mathrm{\beta^-}$ decay with a laboratory half-life of 99.8 yr producing $\mathrm{^{63}Cu}$. According to NETGEN, the $\mathrm{\beta^-}$-decay rate of $\mathrm{^{63}Ni}$ increases by roughly an order of magnitude from the terrestrial value under the conditions that prevail in the intershell (see Table \ref{table:table 3}). As shown in Eq. \ref{eq:1}, the probability of neutron capture decreases with increasing decay rates, therefore, we obtain a lower $\mathrm{^{64}Ni}$ abundance from this channel in Set 1 models than in Set 0 models. The isotope $\mathrm{^{64}Cu}$, with a half-life of 12.8 hours, can both $\mathrm{\beta^-}$ and $\mathrm{\beta^+}$ decay, where the latter channel is twice as fast compared to the $\mathrm{\beta^-}$ under terrestrial conditions. While the rate of the $\mathrm{\beta^-}$ channel of $\mathrm{^{64}Cu}$ slightly increases with \textit{T}, the $\mathrm{\beta^+}$-decay rate is instead reduced by about a third at 300 MK. The probability of the $\mathrm{\beta^+}$ channel also decreases with electron number density. These effects also contribute to the decrease in the amount of $\mathrm{^{64}Ni}$ in Set 1 models. In total, we obtain 1\% to 12\% less $\mathrm{^{64}Ni}$ in the final stage of the Set 1 models compared to Set 0. For nickel, a large number of SiC grain data are available following the work of \citet{tr18}, therefore, we compare the predictions of our models with these data in Section \ref{section:section 3.3.1.}.

\subsubsection{\texorpdfstring{$\mathrm{^{80,81}Kr}$}{}}
\label{section:section 3.1.2.}

The \textit{s}-process production of the stable $\mathrm{^{80}Kr}$ and the unstable $\mathrm{^{81}Kr}$ with a half-life of 229 kyr depends on the operation of the $\mathrm{^{79}Se}$ and $\mathrm{^{80}Br}$ branching points. $\mathrm{^{79}Se}$ suffers $\mathrm{\beta^-}$ decay with a laboratory half-life of 297 kyr producing $\mathrm{^{79}Br}$. Neutron captures on $\mathrm{^{79}Br}$ produce unstable $\mathrm{^{80}Br}$ with a half-life of 17.7 minutes. $\mathrm{^{80}Br}$ can both $\mathrm{\beta^-}$ and $\mathrm{\beta^+}$ decay, the former channel being almost ten times faster than the latter and resulting in the production of $\mathrm{^{80}Kr}$. From $\mathrm{^{80}Kr}$, $\mathrm{^{81}Kr}$ can be produced by neutron captures, in addition to its amount also depending on its own electron capture rate. Comparing the final surface abundances of the Set 1 models with those of Set 0, we obtain 1.94 - 4.65 times more $\mathrm{^{80}Kr}$ and 7.23 - 344.43 times more $\mathrm{^{81}Kr}$. Note that the abundance of $\mathrm{^{82}Kr}$ also increased slightly, by 2-15\%. All these changes are mainly due to the operation of the $\mathrm{^{79}Se}$ branching point, whose decay rate increases by orders of magnitude under stellar conditions (Table \ref{table:table 3}). Note that a measurement-based \textit{T}- and $\rho$-dependent half-life of $\mathrm{^{79}Se}$ is available \citep{kl88}. The half-life from NETGEN is within 10\% of the half-life of \citet{kl88} over the whole temperature range.

\setlength{\tabcolsep}{0.25em}
\begin{table}
\caption{The $\mathrm{^{80}Kr}$/$\mathrm{^{82}Kr}$ ratios in the selected Set 0 and Set 1 models and for each KJ fraction.}
\label{table:4}
\centering
\begin{tabular}{ccccc}
\hline
\hline
\multirow{2}{*}{Model} & \multirow{2}{*}{$\mathrm{\left(^{80}Kr/^{82}Kr \right)_{Set\,0}}$} & \multirow{2}{*}{$\mathrm{\left(^{80}Kr/^{82}Kr \right)_{Set\,1}}$} & \multicolumn{2}{c}{Reference (1)} \\\cline{4-5}
& & & Fraction & $\mathrm{^{80}Kr}$/$\mathrm{^{82}Kr}$ \\
\hline
\hline
\multicolumn{3}{c}{\textit{Z} = 0.014, \textit{Y} = 0.28} & KJA & 0.0511(71) \\
\cline{1-3}
2\,$\mathrm{M_{\odot}}$ & 0.013 & 0.105 & KJB & 0.0500(17) \\
3\,$\mathrm{M_{\odot}}$ & 0.002 & 0.037 & KJC & 0.0436(29) \\
4\,$\mathrm{M_{\odot}}$ & $10^{-6}$ & 0.023 & KJD & 0.0398(23) \\
\cline{1-3}
\multicolumn{3}{c}{\textit{Z} = 0.03, \textit{Y} = 0.30} & KJE & 0.0356(23) \\
\cline{1-3}
3\,$\mathrm{M_{\odot}}$ & 0.007 & 0.079 & KJG & 0.0234(76) \\
4\,$\mathrm{M_{\odot}}$ & 0.001 & 0.040 & KJH & 0.0460(62) \\
\hline
\end{tabular}
\tablebib{(1) \citet{le94}}
\tablefoot{The values represent the corresponding ratios in the intershell of each models before the last TDU episode. The letter after KJ indicates increase in grain size from 0.38 $\mathrm{\mu}$m to 4.57 $\mathrm{\mu}$m.}
\end{table}

Krypton is a noble gas, it does not condense into dust, but it can be ionised and implanted into the grains \citep{ve04}. In fact, the isotopic composition of trace amounts of Kr in SiC grains shows the signature of mixing of pure \textit{s} process and solar composition material. \citet{le94} analysed the isotopic composition of Kr in bulk samples of a million SiC grains from the Murchison KJ series \citep{am94} and used linear regression to derive the isotopic composition of Kr in the intershell of AGB stars. They found that the isotopic ratios vary with grain size (the $\mathrm{^{80}Kr}$/$\mathrm{^{82}Kr}$ ratio decreases with grain size), suggesting that these SiC grains came from stars of different masses and metallicities. 

Table \ref{table:4} shows the $\mathrm{^{80}Kr}$/$\mathrm{^{82}Kr}$ ratios in the intershell of the Set 0 and Set 1 models before the last TDU episode. We compare only the solar and double-solar metallicity model predictions with the \citet{le94} data because
the vast majority (> 90\%) of SiC grains, the mainstream (MS) grains, are believed to originate in low-mass C-rich AGB stars with around solar \citep{ho94}, slightly super-solar \citep{le13} and double-solar \citep{lu20} metallicity. Also recent results by \citet{cr20} suggest that the stardust SiC grains predominantly came from AGB stars around \textit{M} = 2 $\mathrm{M_{\odot}}$ and \textit{Z} = 0.014. For most of the Set 0 models, this ratio is an order of magnitude lower than the range of \citet{le94}, for the \textit{M} = 4 $\mathrm{M_{\odot}}$ \textit{Z} = 0.014 model, this difference is even greater. In contrast, the $\mathrm{^{80}Kr}$/$\mathrm{^{82}Kr}$ ratios derived by the Set 1 models are of the same order of magnitude as the possible intershell ratios, in addition, the predictions of three Set 1 models are within the range observed. Note that a new $\mathrm{^{80}Kr(n,\gamma)}$ cross section is available following the work of \citet{te21}. When testing on \textit{M} = 2 and 3 $\mathrm{M_{\odot}}$ \textit{Z} = 0.014 models, the new neutron-capture rate of $\mathrm{^{80}Kr}$ based on \citet{te21} results in slightly higher (up to 13\%) $\mathrm{^{80}Kr}$/$\mathrm{^{82}Kr}$ ratios and does not change our conclusions above. 
\begin{figure*} [ht!]
	\begin{center}
		\includegraphics[width=\hsize]{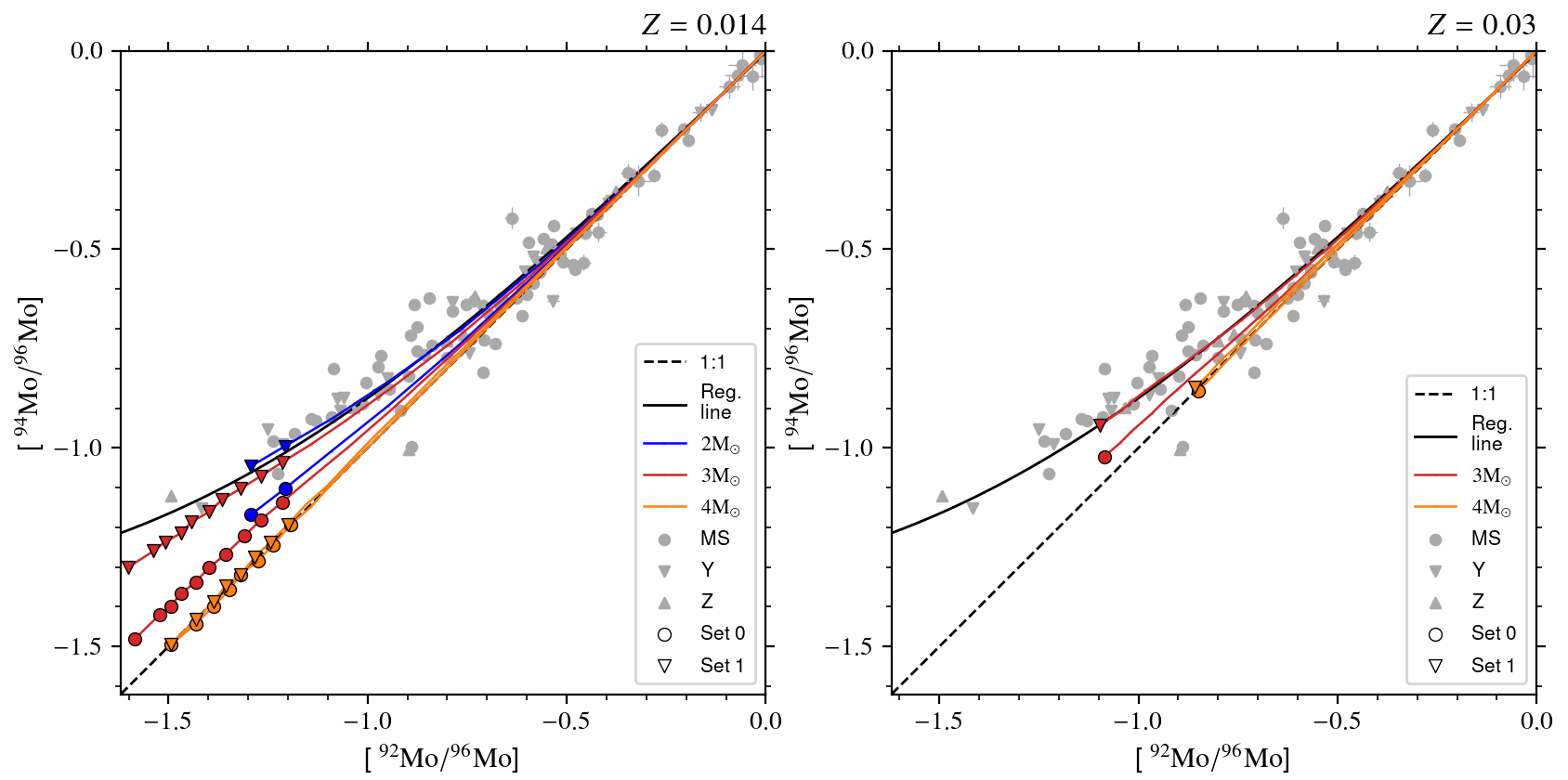}
		\caption{$[\mathrm{^{94}Mo/^{96}Mo}]$ versus $[\mathrm{^{92}Mo/^{96}Mo}]$ from SiC grain data as dark gray circles for mainstream (MS) and down and up triangles for Y and Z grains, respectively with 1$\mathrm{\sigma}$ error bars \citep[from][]{ba07,li17,li19,st19} compared to the surface evolution of Set 0 and Set 1 stellar models of solar (left panel) and of double-solar metallicity (right panel) of different masses (see legend). Circles (Set 0) and triangles (Set 1) on the lines represent the composition after each TDU after the envelope becomes C-rich. The solid black line represents the linear regression from $\mathrm{\delta^{94}Mo_{96}}$ vs $\mathrm{\delta^{92}Mo_{96}}$, while the dashed black line represents the $[\mathrm{^{94}Mo/^{96}Mo}]$ = $[\mathrm{^{92}Mo/^{96}Mo}]$ line.}
  \label{figure:figure 1}
	\end{center}
\end{figure*} 

\subsubsection{\texorpdfstring{$\mathrm{^{94}Mo}$}{}}
\label{section:section 3.1.3.}

By definition, stable proton-rich isotopes that cannot be reached by neutron-capture processes are \textit{p}-only isotopes \citep[or the \textit{p}-nuclei,][]{ca57}. The \textit{p}-only $\mathrm{^{94}Mo}$ can also be produced via the \textit{s} process by two consecutive neutron captures on $\mathrm{^{92}Mo}$ and the operation of $\mathrm{^{93}Zr}$ and $\mathrm{^{94}Nb}$ branching points. $\mathrm{^{94}Mo}$ is the daughter nucleus of $\mathrm{^{94}Nb}$. The latter can be produced by neutron captures on both the initial amount of stable $\mathrm{^{93}Nb}$ and the small fraction of $\mathrm{^{93}Nb}$ that comes from the $\mathrm{\beta^-}$-decay of $\mathrm{^{93}Zr}$. $\mathrm{^{92}Mo}$ is also a \textit{p}-nucleus and it cannot be produced by the \textit{s} process under any circumstances. The $\mathrm{^{92,93}Mo}$ isotopes are not included in our network, therefore, the $\mathrm{^{92}Mo(n,\gamma)^{93}Mo(n,\gamma)^{94}Mo}$ channel is also missing. However, we test that its impact on the abundance of $\mathrm{^{94}Mo}$ is typically less than $\sim$1\%. Although the decay rate of $\mathrm{^{93}Zr}$ increases slightly with temperature and electron density, the decay rate of $\mathrm{^{94}Nb}$ is six orders of magnitude higher in the intershell region of AGB stars than under terrestrial conditions (see Table \ref{table:table 3}). Therefore, while more $\mathrm{^{94}Nb}$ is produced in the Set 1 models compared to Set 0, its abundance also decays much faster, leading to a decrease in the final surface abundance by orders of magnitude and a milder increase (at most 53\%) in the abundance of $\mathrm{^{94}Mo}$. 

Several publications have analysed the isotopic compositions of molybdenum in SiC grains \citep[e.g.][]{ni98,lu03,st19}. As shown in these previous studies, the isotope ratios observed in SiC grains show the mixtures of pure \textit{s}-process matter with solar material. In addition, the $\mathrm{\delta^{94}Mo_{96}}$ values are correlated with the $\mathrm{\delta^{92}Mo_{96}}$ values, and the slope of the regression line indicates the presence of a small \textit{s}-process contribution to $\mathrm{^{94}Mo}$. We calculate the linear regression\footnote{Regression was performed using \textit{scipy.odr} \citep{scipy} and the uncertainties of the data points were taken into account.} of $\mathrm{\delta}$ values from \citet{ba07}, \citet{li17,li19} and \citet{st19} and we obtain an intercept of $\mathrm{\delta^{94}Mo_{96}}$ = -961.8\textperthousand\; $\pm$ 1.2\textperthousand\; at $\mathrm{\delta^{92}Mo_{96}}$ = -1000\textperthousand\; which is consistent with the result of \citet{st19}. We can estimate the \textit{s}-process contribution $s_{\mathrm{i}}$ for $\mathrm{^{94}Mo}$ from our models using the following equation:
\begin{equation}
\label{eq:6}
s_{\mathrm{i}} =\left ( \frac{M^{\mathrm{wm}}_{\mathrm{i}}\,X^{\mathrm{^{150}Sm}}_{\mathrm{\odot}}}{X^{i}_{\mathrm{\odot}}\,M^{\mathrm{wm}}_{\mathrm{^{150}Sm}}} \right ),
\end{equation}
where $M^{\mathrm{wm}}_{\mathrm{i}}$ is the weighted mean yield (see below) of isotope $i$ obtained by our stellar models normalised to the unbranched \textit{s}-only isotope $\mathrm{^{150}Sm}$, whose solar abundance and neutron capture cross section are known with high precision \citep{ga23}. The weighted mean yields are calculated with a trapezoidal rule quadrature where, with a constant mass interval, simplifies to: 
\begin{equation}
\label{eq:7}
M^{\mathrm{wm}}_{\mathrm{i}} = \frac{\sum_{\mathrm{j}=0}^{\mathrm{n}-1} \left ( M_{\mathrm{i,j}}M^{-2.35}_{\mathrm{j}} + M_{\mathrm{i,j+1}}M^{-2.35}_{\mathrm{j+1}} \right )}{\sum_{\mathrm{j}=0}^{\mathrm{n}-1} \left (M^{-2.35}_{\mathrm{j}} + M^{-2.35}_{\mathrm{j+1}} \right )},
\end{equation}
where $j$ indicates the mass in question. The weight is based on the Salpeter initial mass function \citep[IMF,][]{sa55}, used here to mimic the yields for the entire AGB stellar population. Our prediction to the \textit{s}-process contribution of $\mathrm{^{94}Mo}$ does not change significantly: it slightly increases from 0.034 to 0.04 from Set 0 to Set 1, and both are consistent with $\sim$0.038 coming from the data regression \citep[see also][]{st19}. 

In addition, we can compare the surface evolution of each model with the grain data. For easier comparison to model predictions, in Fig. \ref{figure:figure 1} we plot the logarithms of the ratio of the isotopes $i$ and $j$ at each thermal pulse $k$ normalised to solar composition using standard astrophysical notation:
\begin{equation}
[\mathrm{^{i}E/^{j}E]} = \mathrm{log(}Y^{\mathrm{i}}/Y^{\mathrm{j}}\mathrm{)}_{\mathrm{sample}} - \mathrm{log(}Y^{\mathrm{i}}/Y^{\mathrm{j}}\mathrm{)}_{\mathrm{\odot}}.
\end{equation}
As in the case of $\mathrm{^{80}Kr}$, we only use the predictions of the solar and double-solar metallicity models, but include the few data points available for grains from the minor Y and Z populations, believed to have originated in stars of sub-solar metallicity, because in Fig. \ref{figure:figure 1} these grains plot in the same region as the mainstream grains. The two 4 $\mathrm{M_{\odot}}$ models are on the 1:1 line regardless of the choice of the reaction rates, which means that the $\mathrm{^{94}Mo}$ production is not relevant in these stars, therefore, they do not represent the majority of the grain parent stars, especially in the case of the solar metallicity model \citep{lu18a}. On the other hand, the surface evolution of the 2 and 3 $\mathrm{M_{\odot}}$ Set 1 models closely follows the regression line. Most previous models \citep[e.g.][]{ar99,lu03,bi14,li19} were unable to produce enough $\mathrm{^{94}Mo}$ to match the grain data, our new models resolve this problem. Models in FRUITY \citep[FUll-Network Repository of Updated Isotopic Tables \& Yields, e.g.][]{cr09,cr11,cr15} also produce a significant amount of $\mathrm{^{94}Mo}$, actually resulting in values lying even above the line defined by the grains \citep{ve20}. Therefore, we conclude that the previous mismatch was likely due to inaccurate nuclear input data and/or treatment of the branching points, which was updated in the past decade or so in the new generation of AGB nucleosynthesis models. 

\subsubsection{\texorpdfstring{$\mathrm{^{108}Cd}$}{}}
\label{section:section 3.1.4.}

The case of $\mathrm{^{108}Cd}$ is similar to $\mathrm{^{94}Mo}$: it is a \textit{p}-nucleus with a small \textit{s}-process contribution. It can be produced via \textit{s} process by the $\mathrm{\beta^-}$-decay of $\mathrm{^{108}Ag}$, which can be produced by neutron captures on the initial amount of $\mathrm{^{107}Ag}$ and the small fraction that comes from the $\mathrm{\beta^-}$-decay of $\mathrm{^{107}Pd}$. $\mathrm{^{108}Ag}$ can both $\mathrm{\beta^-}$ and $\mathrm{\beta^+}$ decay with a relatively short half-life (2.4 min). The $\mathrm{\beta^-}$ channel does not depend significantly on temperature and density and is faster than the $\mathrm{\beta^+}$ channel under both stellar and terrestrial conditions (see Table \ref{table:table 3}). In contrast, the $\mathrm{\beta^-}$-rate of $\mathrm{^{107}Pd}$ increases by orders of magnitude with temperature, causing the amount of $\mathrm{^{108}Cd}$ to increase by 2-64\% in the Set 1 models with respect to the Set 0 models. We estimate the \textit{s}-process contribution for $\mathrm{^{108}Cd}$ according to the method used for $\mathrm{^{94}Mo}$. As in the case of $\mathrm{^{94}Mo}$, our predictions of the two sets are quite close: the estimated \textit{s}-process contribution of the solar $\mathrm{^{108}Cd}$ abundance increases slightly from 0.04 to 0.048 from Set 0 to Set 1. 

\subsubsection{\texorpdfstring{$\mathrm{^{129}I}$}{}}
\label{section:section 3.1.5.}

The small \textit{s}-process production of \textit{r}-only $\mathrm{^{129}I}$ depends on the $\mathrm{\beta^-}$ decay of $\mathrm{^{128}I}$. By definition, species that cannot be produced by the \textit{s} process due to an unstable nucleus that precedes them on the \textit{s}-process path are traditionally \textit{r}-only isotopes. In fact, no isotope is completely shielded against \textit{s}-process production, nuclei on the neutron-rich side of the $\mathrm{\beta}$-stability valley can have a low \textit{s} contribution, depending on the activation of branching points \citep{bi15}. In addition, $\mathrm{^{129}I}$ is a so-called short-lived radionuclide \citep[SLR, a radioactive isotope with a half-life between about 0.1 to 100 million years, see, e.g.][]{lu18b}, therefore, its amount depends also on its decay. Neither the decay of $\mathrm{^{128}I}$ nor $\mathrm{^{129}I}$ is significantly dependent on the electron density, but the $\mathrm{\beta^-}$-decay rate of $\mathrm{^{128}I}$ decreases slightly with temperature, while the decay rate of $\mathrm{^{129}I}$ increases by orders of magnitude with temperature (see Table \ref{table:table 3}). While we produce more $\mathrm{^{129}I}$ from this channel in the Set 1 models than in the Set 0 models, its final amount decreases by up to an order of magnitude due to its shorter half-life at stellar temperatures. 

The presence of SLRs has been documented in the early Solar System (ESS) from the laboratory analysis of meteorites \citep[e.g.][]{da11,da22}. These nuclei have now decayed completely, but their ESS abundances can be inferred from measurable excesses in the abundances of their daughter nuclei. Therefore, we cannot determine the solar \textit{s}-contribution, but we can calculate the so-called abundance production factor $P_{\mathrm{s}}$, that is, the relative amount produced by the \textit{s} process compared its stable reference isotope:
\begin{equation}
\label{eq:9}
P_{\mathrm{s}} = \frac{M^{\mathrm{wm}}_{\mathrm{i}}\,m_{\mathrm{j}}}{M^{\mathrm{wm}}_{\mathrm{j}}\,m_{\mathrm{i}}}, 
\end{equation}
where $i$ represents the radioactive and $j$ represents the stable isotope of the corresponding element. The $\mathrm{^{129}I}$/$\mathrm{^{127}I}$ production factors derived by our Set 0 and Set 1 models are 1 $\times$ $10^{-4}$ and 2.8 $\times$ $10^{-4}$, respectively. These values are similar to the previous \textit{Monash} result \citep[$P_{\mathrm{s}}$ $\simeq$ $10^{-4}$,][]{lu14a}, based on AGB models with initial mass between 1.25 and 8.5 $\mathrm{M_{\odot}}$. These production factors are generally four magnitudes lower than the \textit{r}-process production factor $P(r)$ = 1.35 \citep{lu14a,co21}, which confirms that the $\mathrm{^{129}I}$ contribution from the \textit{s}-process is very marginal. 

\subsubsection{\texorpdfstring{$\mathrm{^{152}Gd}$}{}}
\label{section:section 3.1.6.}

The \textit{s}-process production of $\mathrm{^{152}Gd}$ depends on the operation of the $\mathrm{^{151}Sm}$ and $\mathrm{^{152}Eu}$ branching points. $\mathrm{^{152}Eu}$ can capture electrons and undergo $\mathrm{\beta^-}$ decay with a laboratory half-life of 13.3 yr. The rate of electron capture increases with both temperature and electron density, while the rate of the $\mathrm{\beta^-}$ channel increases only with temperature. The amount of $\mathrm{^{152}Eu}$ produced is determined by the $\mathrm{\beta^-}$-decay of $\mathrm{^{151}Sm}$, whose rate also increases with temperature depending on the electron density (see Table \ref{table:table 3}). Comparing the Set 1 models to the Set 0 models, we produce more $\mathrm{^{152}Eu}$, which is more likely to suffer $\mathrm{\beta^-}$ decay under stellar conditions than under terrestrial conditions, therefore, we obtain $\sim$2-28 times more $\mathrm{^{152}Gd}$, depending on the stellar model.

$\mathrm{^{152}Gd}$ was one of the 35 original \textit{p}-nuclei \citep{ca57}, but nowadays it is well accepted that it has a large \textit{s}-process contribution \citep{ar99,bi11,bi14}, or it is even considered as an \textit{s}-only isotope \citep{tr11}. \citet{ar99} calculated \textit{s}-process contributions using the arithmetic average yields of 1.5 and 3 $\mathrm{M_{\odot}}$ half-solar metallicity AGB models from \citet{ga98} and normalised to solar abundances by \citet{an89}. \citet{bi11} updated their results, taking into account more recent neutron-capture cross sections and using solar abundances from \citet{lo09} to normalise. The \textit{s}-contribution for $\mathrm{^{152}Gd}$ is 0.883 and 0.705 in \citet{ar99} and \citet{bi11}, respectively. \citet{bi14} obtained the \textit{s}-components of the solar composition using the GCE code described by \citet{tr04} and \textit{s}-process yields from AGB stellar models by \citet{ga98}. They predicted 0.865 for the \textit{s}-contribution of $\mathrm{^{152}Gd}$ from low-mass AGB stars. We use the method described in Sect. \ref{section:section 3.1.3.} to obtain the \textit{s}-contributions of $\mathrm{^{152}Gd}$. Comparing the predictions of the Set 0 and Set 1 models shows that we need to accurately model the operation of the branching points to understand the nucleosynthesis origin of this isotope. In the Set 0 models, the solar \textit{s}-contributions of $\mathrm{^{152}Gd}$ is only 0.072, which would lead us to conclude that $\mathrm{^{152}Gd}$ is produced mainly by the \textit{p} process. In contrast, Set 1 models predict 0.658 solar \textit{s}-contributions, which confirm that this isotope predominantly originates from low-mass AGB stars. This prediction is broadly consistent with recent results \citep{pr18,bu21} and with the results presented above, especially with \citet{bi11}, whose method to calculate the \textit{s}-process contributions was similar to ours, but used only two models. We emphasise that our method gives an approximate estimate, while the \citet{bi14} results were based on a full GCE model, which is a more accurate approach because it includes many more models also of much lower metallicity than considered here, and evolves each star according to its mass and therefore lifetime.

\subsubsection{\texorpdfstring{$\mathrm{^{176}Lu}$ and $\mathrm{^{176}Hf}$}{}}
\label{section:section 3.1.7.}

The $\mathrm{\beta^-}$-decay rate of $\mathrm{^{176}Lu}$ is highly sensitive to temperature during thermal pulses \citep{ta87,kl91,do99} due to the coupling of the very long-lived ground state ($t_{\mathrm{1/2}}$ = 37.9 Gyr) and the short-lived isomer ($t_{\mathrm{1/2}}$ = 3.7 h) at 123 keV via a thermal population of mediating states at higher energies, in particular at $E$ = 838.6 keV, since direct transition is forbidden by selection rules. The high \textit{T} dependence affects the abundance ratio of $\mathrm{^{176}Lu}$ and its daughter nucleus $\mathrm{^{176}Hf}$ in AGB stars, which must match the relative solar abundances of these isotopes, since both isotopes are \textit{s}-only, shielded by $\mathrm{^{176}Yb}$ against \textit{r}-process production. In other words, these isotopes must have the same overabundances as $\mathrm{^{150}Sm}$, that is, their \textit{s}-contributions must be close to 1, a minor deviation in the \textit{s}-contribution (1.05 $\pm$ 0.05 for $\mathrm{^{176}Lu}$ and 1 $\pm$ 0.05 for $\mathrm{^{176}Hf}$) is acceptable \citep{he08a} due to the $\mathrm{\beta^-}$ decay of $\mathrm{^{176}Lu}$ in the ISM.

The discrepancy between the theoretical and observed $\mathrm{^{176}Hf}$/$\mathrm{^{176}Lu}$ ratio is a long-standing problem that has been addressed in a number of studies over the past decades \citep[e.g.][]{he08a,mo09,cr10}. According to the description of \citet{kl91}, the branching factor of $\mathrm{^{176}Lu}$ is a complex function of temperature and neutron density and depends, in particular, on the ratio of the two partial (n,$\gamma$) cross sections of $\mathrm{^{175}Lu}$ to the ground state and to the isomer state of $\mathrm{^{176}Lu}$, the lifetime of the first mediating state and the decay branching of the first mediating state toward the isomer. By branching analysis, \citet{he08a} found that a plausible value of the isomeric to total ratio IR of 0.81-0.82 resulted in a $\mathrm{^{176}Hf}$/$\mathrm{^{176}Lu}$ ratio consistent with the solar abundances. Less plausible values of 0.8 - 0.75 implemented in a large set of AGB models \citep{cr09} and then in a full GCE model \citep{pr20} provided result within 4$\sigma$ of the solar abundances. However, new experimental data on the nuclear states of $\mathrm{^{176}Lu}$ \citep{mo09} led to an overproduction of $\mathrm{^{176}Lu}$, which was later confirmed by \citet{cr10}. Current ideas suggest that the solution should be sought in the coupling scheme between the thermally populated states of $\mathrm{^{176}Lu}$ \citep[][]{cr10,bi15}. 

We calculate the solar \textit{s}-process contribution for $\mathrm{^{176}Lu}$ and $\mathrm{^{176}Hf}$ by Eq. \ref{eq:6} and \ref{eq:7} with Set 1 models, and we strongly overproduce $\mathrm{^{176}Lu}$ and underproduce $\mathrm{^{176}Hf}$: their \textit{s}-contributions are 2.36 and 0.56, respectively. This is because the NETGEN rate of $\mathrm{^{176}Lu}$ assumes that the ground state and isomer are in thermal equilibrium under \textit{s}-process conditions, but this is only true above 300 MK \citep[e.g.][]{he08a}. Interestingly, our results are similar to those obtained by \citet[][2.4 and 0.55 for $\mathrm{^{176}Lu}$ and $\mathrm{^{176}Hf}$, respectively]{cr10} using a 2 $\mathrm{M_{\odot}}$ \textit{Z} = 0.02 FRANEC model and the $\mathrm{\beta^-}$-decay rate of $\mathrm{^{176}Lu}$ based on parameters of \citet{mo09}. 

\setlength{\tabcolsep}{1.6em}
\begin{table}
\caption{The $\mathrm{^{182}Hf}$/$\mathrm{^{180}Hf}$ production factors of selected Set 0 and Set 1 models, alongside the values previously used in GCE models.}
\label{table:5}
\centering
\begin{tabular}{cccc}
\hline
\hline
Model & \multicolumn{3}{c}{$\mathrm{^{182}Hf}$/$\mathrm{^{180}Hf}$ production factor} \\\cline{2-4}
& Set 0 & Set 1 & Reference (1) \\
\hline
\multicolumn{4}{c}{\textit{Z} = 0.007, \textit{Y} = 0.26} \\ 
\hline
3\,$\mathrm{M_{\odot}}$ & 0.22 & 0.21 & 0.23 \\
\hline
\multicolumn{4}{c}{\textit{Z} = 0.014, \textit{Y} = 0.28} \\ 
\hline
2\,$\mathrm{M_{\odot}}$ & 0.03 & 0.02 & 0.03 \\
3\,$\mathrm{M_{\odot}}$ & 0.10 & 0.09 & 0.12 \\
4\,$\mathrm{M_{\odot}}$ & 0.29 & 0.28 & 0.34 \\
\hline
\multicolumn{4}{c}{\textit{Z} = 0.03, \textit{Y} = 0.30} \\
\hline
3\,$\mathrm{M_{\odot}}$ & 0.03 & 0.02 & 0.04 \\
4\,$\mathrm{M_{\odot}}$ & 0.11 & 0.10 & 0.18 \\
\hline
\end{tabular}
\tablebib{(1) \citet{tr22}}
\end{table}

\subsubsection{\texorpdfstring{$\mathrm{^{182}Hf}$}{}}
\label{section:section 3.1.8.}

The \textit{s}-process production of $\mathrm{^{182}Hf}$ depends on the $\mathrm{^{181}Hf}$ branching point. The $\mathrm{\beta^-}$-decay rate of $\mathrm{^{181}Hf}$ is based on \citet{lu14a} and was already included in the nuclear network of Set 0. In addition to this, the amount of $\mathrm{^{182}Hf}$ depends on its own decay. $\mathrm{^{182}Hf}$ is a short-lived radionuclide with a terrestrial half-life of 8.8 Myr. Its $\mathrm{\beta^-}$-decay rate is \textit{T}- and $\rho$-dependent, and it decreases by six orders of magnitude under stellar conditions, according to NETGEN. We obtain 5\%-28\% less $\mathrm{^{182}Hf}$ on the surface at the final stage of the Set 1 models compared to Set 0. 

\begin{figure*} [ht!]
	\begin{center}
		\includegraphics[width=\hsize]{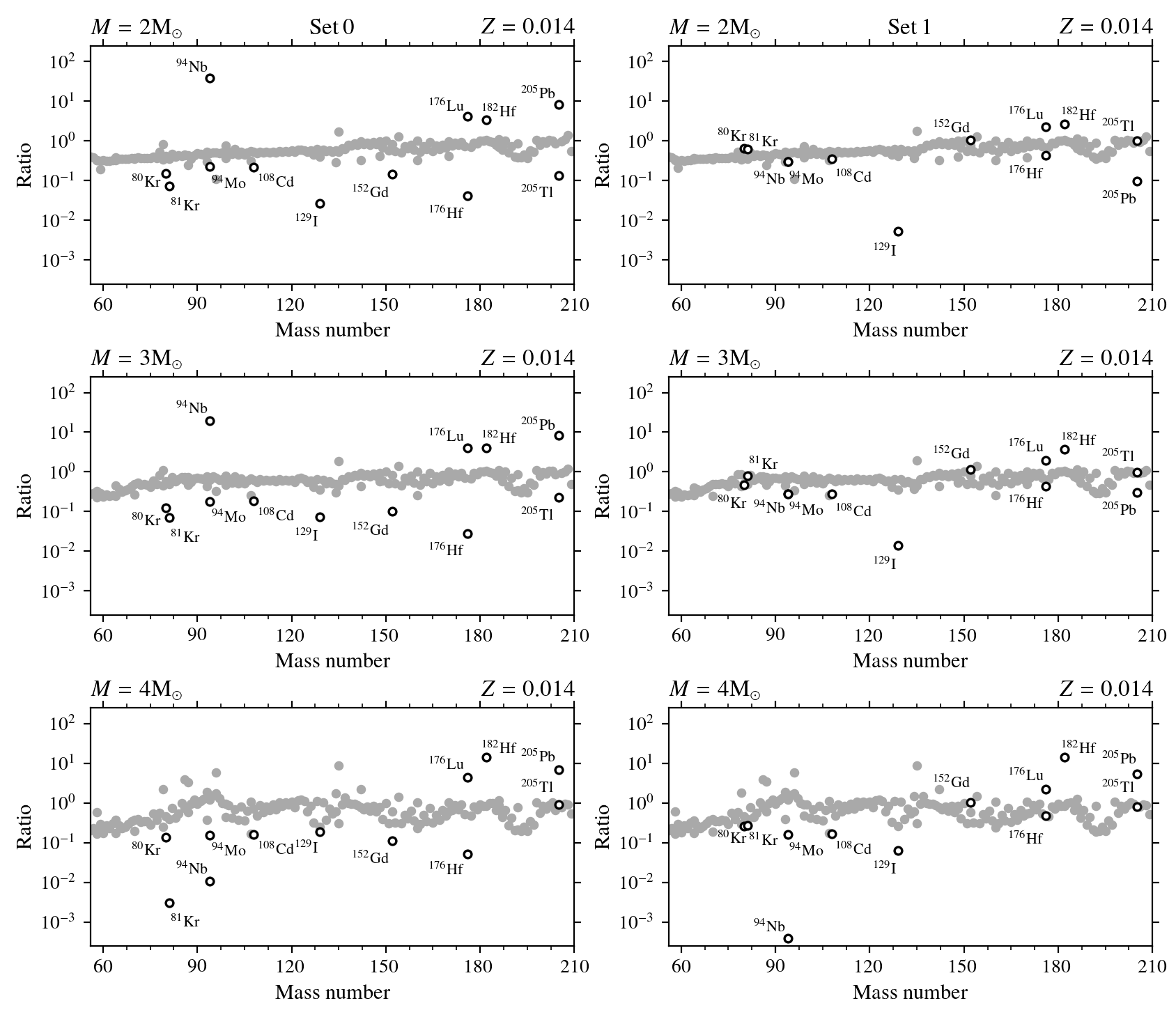}
		\caption{Ratio between the final surface isotopic distributions (normalised to the abundance of $\mathrm{^{150}Sm}$) of the Set 0 (left panel) and Set 1 (right panel) with respect to the FRUITY calculations of solar metallicity models of different masses, as indicated. }
      \label{figure:figure 2}
	\end{center}
\end{figure*} 

As in the case of $\mathrm{^{129}I}$, we can calculate the production factor of $\mathrm{^{182}Hf}$. We can compare the $\mathrm{^{182}Hf}$/$\mathrm{^{180}Hf}$ production factors derived by individual stellar models with the \textit{Monash} production factors used by \citet[][Table 6]{tr22}. These authors used two sets of AGB yields and calculated timescales relevant for the birth of the Sun using the OMEGA+ GCE code \citep{co18} and SLRs heavier than iron produced by the \textit{s} process. The production factors of \citet{tr22} are slightly higher than our Set 0 results (see Table \ref{table:5}), because their \textit{Monash} yields came from \citet{ka16}, whose models include the terrestrial $\mathrm{\beta^-}$-decay rate of $\mathrm{^{181}Hf}$ without any \textit{T}- and $\rho$-dependence, which is slightly higher in the stellar interior than under terrestrial conditions. A further small decrease in production factors is also present between the Set 0 and Set 1 models, as expected from the introduction of the $\mathrm{^{182}Hf}$ \textit{T}-dependent decay.

\subsubsection{\texorpdfstring{$\mathrm{^{205}Pb}$}{}}
\label{section:section 3.1.9.}

In the case of $\mathrm{^{205}Tl}$ and $\mathrm{^{205}Pb}$, two processes compete with each other depending on temperature and electron density: the bound-state $\mathrm{\beta}$ decay of $\mathrm{^{205}Tl}$ and the electron capture of $\mathrm{^{205}Pb}$. While $\mathrm{^{205}Tl}$ is stable under terrestrial conditions, its bound-state decay is active only under stellar conditions, and its rate increases with temperature \citep{le24}. In contrast, the rate of electron capture of $\mathrm{^{205}Pb}$ strongly increases with temperature up to 50-100 MK (depending on the electron number density), and then it decreases slightly. According to NETGEN, the rates of these two processes are equal somewhere between 130 MK and 310 MK, the exact temperature of the intersection increasing with increasing electron number density. This means that the $\mathrm{^{205}Tl}$/$\mathrm{^{205}Pb}$ ratio in stellar models strongly depends on the temperature condition (e.g. maximum temperature, average temperature, and their time evolution) of the intershell. We produce less $\mathrm{^{205}Pb}$ and more $\mathrm{^{205}Tl}$ in the Set 1 models compared to Set 0 models because the electron-capture reaction is stronger than the bound-state decay except for the 4 $\mathrm{M_{\odot}}$ \textit{Z} = 0.007 model, which is hot enough to overproduce $\mathrm{^{205}Pb}$, that is, more $\mathrm{^{205}Tl}$ decays to $\mathrm{^{205}Pb}$ than vice versa, and the 4 $\mathrm{M_{\odot}}$ \textit{Z} = 0.014 model, where the abundances of both isotopes decrease relative to the Set 0 case, because less $\mathrm{^{205}Tl}$ is produced from the $\mathrm{^{204}Tl}(n,\gamma)\mathrm{^{205}Tl}$ channel. $\mathrm{^{204}Tl}$ is a branching point, it suffers $\mathrm{\beta^-}$-decay with a terrestrial half-life of 3.9 yr, which is much shorter in stellar interior (Table \ref{table:table 3}).

Using the SLRs and their abundance production factors, we can obtain information on the chronology of the evolution of the ESS, including the time interval between the separation of Solar System material from ISM and the formation of the first solids. This so-called "isolation time" corresponds to the free decay time required for the abundance ratio predicted in the ISM to reach the ESS ratio measured in meteorites \citep{lu18b}. $\mathrm{^{205}Pb}$ is a unique isotope because among the measurable SLRs heavier than iron, it is the only that cannot be produced by the \textit{r} process. In addition, its reference isotope $\mathrm{^{204}Pb}$ is also a \textit{s}-only. Despite this, it has not been extensively used in the literature for calculating the isolation time of the Solar System, because of both the high uncertainty of its ESS ratio \citep{pa18} and the strongly \textit{T}- and $\rho$-dependent electron-capture rate, which is not well determined \citep{mo98}. We attempt to calculate the isolation time using the $\mathrm{^{205}Pb}/\mathrm{^{204}Pb}$ ratio as
\begin{equation}
T_{\mathrm{iso}} = - \mathrm{ln} \left [ \frac{\left ( \mathrm{^{205}Pb}/\mathrm{^{204}Pb} \right )_{\mathrm{ESS}}}{\left ( \mathrm{^{205}Pb}/\mathrm{^{204}Pb} \right )_{\mathrm{ISM}}}\right]\tau_{\mathrm{^{205}Pb}},
\end{equation}
where $\tau_{\mathrm{^{205}Pb}}$ = 24.5 Myr is the mean life of $\mathrm{^{205}Pb}$ \citep[value is taken from][]{ko21}, $\left (\mathrm{^{205}Pb}/\mathrm{^{204}Pb} \right )_{\mathrm{ESS}}$ = (1.8 $\pm$ 0.6) $\times$ $10^{-3}$ \citep[value is taken from][with 1$\sigma$ error bar]{pa18} and $\left ( \mathrm{^{205}Pb}/\mathrm{^{204}Pb} \right )_{\mathrm{ISM}}$ are the ESS and ISM ratio of $\mathrm{^{205}Pb}$ to $\mathrm{^{204}Pb}$, respectively. The ISM ratio can be estimated from the following steady-state equation: 

\begin{equation}
\label{eq:11}
\left ( \mathrm{^{205}Pb}/\mathrm{^{204}Pb} \right )_{\mathrm{ISM}} = P \times K \times \frac{\tau_{\mathrm{^{205}Pb}}}{T_{\mathrm{Gal}}},
\end{equation}

where $P$ is the stellar production factor, $T_{\mathrm{Gal}}$ = $8.4$ Gyr is the galactic age at the formation of Solar System and $K$ = 1.6 (min), 2.3 (best) or 5.7 (max), is the GCE parameter that represents the uncertainties coming from galactic evolution \citep{co19}. The obtained ISM ratios in the unit of $10^{-4}$ are 2.4, 3.4 and 8.5, for $K$ = 1.6, 2.3 and 5.7, respectively. 

These ISM ratios are lower than the ESS ratio, therefore, we cannot obtain physical, that is, positive isolation times. Note that we used Eq. \ref{eq:11} with only one production factor calculated from Eq. \ref{eq:7} and \ref{eq:9} from solar metallicity models, while the GCE model of \citet{tr22} handles the production factors of each stellar model. In other words, we use approximate equations, therefore, our method is not as accurate as a full GCE model. However, it is very unlikely that the full GCE models could solve this problem because it has been shown that Eq. \ref{eq:11} reproduces the full GCE models within 50\% \citep{le24}, therefore, we could obtain a value just above the minimum ESS value (allowed within 1$\sigma$) only by applying the maximum value of $K$ and the maximum +50\% variation.

This long-standing problem of $\mathrm{^{205}Pb}$ was recently solved by \citet{le24}, who present new measurement-based rates of the bound-state $\mathrm{\beta}$ decay of $\mathrm{^{205}Tl}$ and the electron capture of $\mathrm{^{205}Pb}$ for a wide range of astrophysical conditions. Using these new rates and a method similar to that described in this work, these authors were able to derive positive isolation times that are consistent with other \textit{s}-process chronometers. 

\subsection{Comparison of Set 1 to FRUITY results}
\label{section:section 3.2.}

With the updated $\mathrm{\beta}$-decay rates, the isotopic predictions of our models can be compared to those of the FRUITY models, as these were calculated including the \textit{T}- and $\rho$-dependence. Because the \textit{Z} = 0.007 and \textit{Z} = 0.03 metallicity models are not available in the FRUITY database, we use the \textit{Z} = 0.014 models for comparison. In general, using the NETGEN rates, most of the abundances of the isotopes discussed in Sect. \ref{section:section 3.1.} are closer to the FRUITY predictions, as expected. Figure \ref{figure:figure 2} shows the ratio of the final surface abundances normalised to the abundance of $\mathrm{^{150}Sm}$ of Set 0 and Set 1 relative to FRUITY for isotopes with a mass number greater than \textit{A} = 56. The highlighted isotopes are those that showed the highest differences between the Set 0 and Set 1 models and were discussed in Sect. \ref{section:section 3.1.}. 

The remaining differences between predictions from our models and FRUITY can result from: (1) Somewhat different nuclear inputs. In fact, FRUITY's temperature and density-dependent $\mathrm{\beta}$-decay rates of isotopes heavier than iron are based on \citet{ta87} with a few exceptions \citep[see][for details]{st06} and use different functions than NETGEN, and therefore \textit{Monash} code to interpolate from the lowest temperature available in the database to the terrestrial values (log versus linear, respectively). (2) In our post-processing code, the equations that represent the abundance changes include nuclear burning and mixing, whereas the version of the FRANEC code that was used for the FRUITY database combines the calculation of stellar structure and nucleosynthesis, but not mixing. (3) Our method of including the $\mathrm{^{13}C}$ pocket differs from that used to produce the FRUITY results, which is based on the time-dependent convective overshoot at the base of the envelope. In addition, the $\mathrm{^{13}C}$ pocket is kept constant in our models, while it self-consistently decreases with pulse number in the FRUITY models. (4) Because of different methods and input physics, the total number of thermal pulses differs, our models have more TPs (21, 26 and 22 vs 13, 17 and 9, for 2, 3 and 4 $\mathrm{M_{\odot}}$ models, respectively). In addition, the maximum temperatures reached by the thermal pulses are also different: while our 2 $\mathrm{M_{\odot}}$ \textit{Z} = 0.014 model is cooler than the FRUITY model (280 MK versus 291 MK), 3 $\mathrm{M_{\odot}}$ \textit{Z} = 0.014 models reach quite similar maximum temperatures (302 MK versus 304 MK), our 4 $\mathrm{M_{\odot}}$ \textit{Z} = 0.014 model reaches higher maximum \textit{T} (348 MK versus 312 MK) than the FRUITY stars. The effect of temperature difference is most visible at 4 $\mathrm{M_{\odot}}$, where the branching points are more active and cause a larger fluctuation in the abundances. 

\begin{table}
\caption{Ratio of MACSs used in Set 2 and Set 1 for the isotopes discussed in Sect. \ref{section:section 3.3.}.}
\setlength{\tabcolsep}{0.95em}
\label{table:6}
\centering
\begin{tabular}{cccccc}
\hline
\hline
& \multicolumn{5}{c}{$\mathrm{ \left( MACS \right)_{Set\,2}}$/$\mathrm{ \left( MACS \right)_{Set\,1}}$ } \\\cline{2-6}
Energy (keV) & 10 &	15 & 20	& 25 & 30 \\
\hline
$\mathrm{^{64}Ni(n,\gamma)}$ & 1.59	& 1.29 & 1.11 & 0.98 & 0.89 \\
$\mathrm{^{134}Ba(n,\gamma)}$ &	1.12 & 1.10 & 1.08 & 1.06 & 1.05 \\
$\mathrm{^{135}Ba(n,\gamma)}$ &	1.07 & 1.07 & 1.08 & 1.07 & 1.07 \\
$\mathrm{^{136}Ba(n,\gamma)}$ &	1.12 & 1.11 & 1.10 & 1.10 & 1.10 \\
$\mathrm{^{137}Ba(n,\gamma)}$ &	1.44 & 1.30	& 1.24 & 1.21 &	1.18 \\
$\mathrm{^{186}W(n,\gamma)}$ & 0.79	& 0.80 & 0.79 & 0.78 & 0.75 \\
\hline
\end{tabular}
\tablefoot{The values are given in the energy range of E = 10-30 keV. The typical energy for $\mathrm{^{13}C(\alpha,n)^{16}O}$ is $\simeq$ 8 keV, for $\mathrm{^{22}Ne(\alpha,n)^{26}Mg}$ $\simeq$ 23 keV. The Set 1 rates are based on the \textit{ka02} fit from JINA reaclib or came from the KaDoNiS database, while the Set 2 rates are based on the ASTRAL database.}
\end{table}

\subsection{Comparison of Set 2 to Set 1}
\label{section:section 3.3.}

The extent of the difference between the predictions of Set 2 and Set 1 is affected by how much of the given isotope is produced by each model. This in turn is determined by the features of the model, the total number of neutrons available, and the differences in the nuclear input. The total amount of free neutrons is usually expressed by the time-integrated neutron flux, that is, the neutron exposure: 
\begin{equation}
\tau = \int N_\mathrm{n}\,v_{\mathrm{th}}\,\mathrm{dt},
\end{equation}
The neutron exposure depends on neutron sources and the amount of Fe seed nuclei. Of the two neutron-source reactions, $\mathrm{^{13}C(\alpha,n)^{16}O}$ results in a higher neutron exposure than $\mathrm{^{22}Ne(\alpha,n)^{26}Mg}$, because, while only a few percent of $\mathrm{^{22}Ne}$ nuclei burn, all of the $\mathrm{^{13}C}$ nuclei are completely destroyed. While the $\mathrm{^{13}C}$ abundance, thus the number of neutrons produced does not vary with the metallicity, the amount of Fe seeds increases with increasing metallicity. Therefore, a higher Fe abundance leads to a lower neutron exposure, as there are more Fe seeds to capture neutrons and a smaller number of free neutrons remains for the production of the \textit{s}-process isotopes. The neutron exposure determines whether the neutron flux can overcome the bottlenecks at stable neutron magic nuclei (neutron number \textit{N} = 28, 50, 82). Once a bottleneck is bypassed, the neutron-capture flow can reach all isotopes up to the next bottleneck, and the abundances between the bottlenecks are in roughly steady-state equilibrium \citep{lu23}:
\begin{equation}
Y^{\mathrm{i}}\,\langle \sigma v \rangle_\mathrm{i} \simeq \mathrm{constant}.
\end{equation}

The final surface normalised number abundances $Y^i_{final}$ of 34 isotopes changed by at least 10\% in at least one of the Set 2 models with respect to the Set 1 models. The list of these isotopes and their $Y^\mathrm{i}_{\mathrm{Set\,2,final}}$/$Y^\mathrm{i}_{\mathrm{Set\,1,final}}$ ratio are shown in Table \ref{table:A.2.}. In the following, we discuss the most relevant cases and in Table \ref{table:6} we list the ratio of MACSs used in Set 2 and Set 1 for the related isotopes. 

\begin{figure} [ht!]
	\begin{center}
		\includegraphics[width=\hsize]{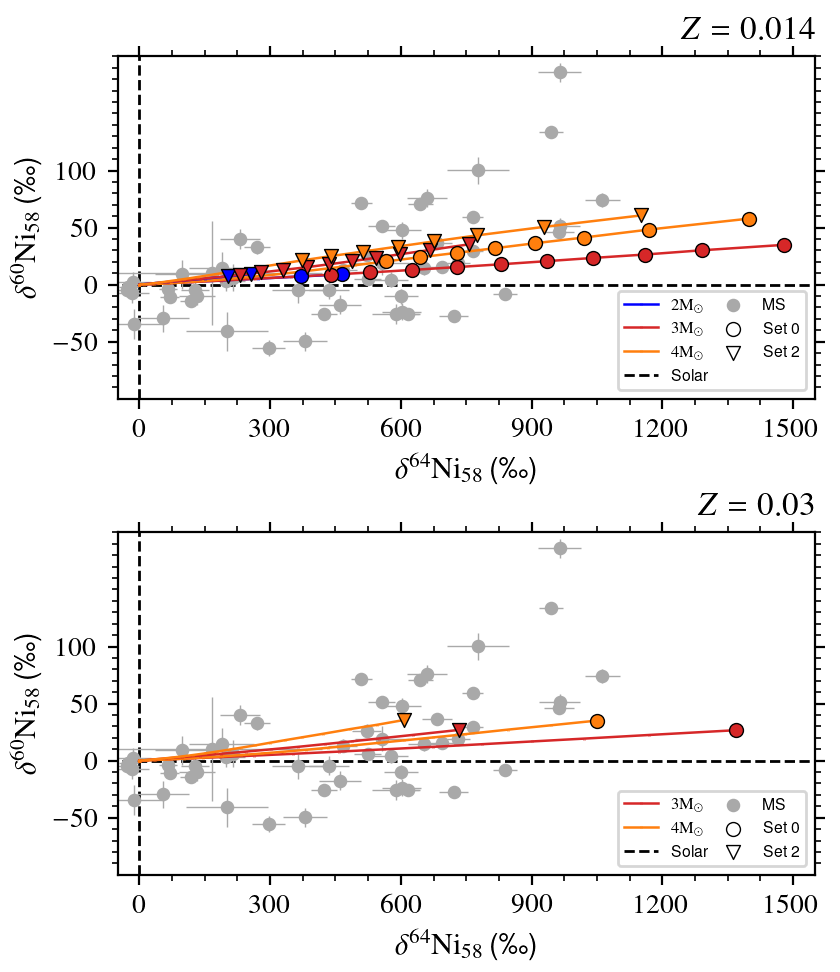}
		\caption{$\mathrm{\delta^{60}Ni_{58}}$ versus $\mathrm{\delta^{64}Ni_{58}}$ from mainstream SiC grain data \citep[dark gray circles with 1$\mathrm{\sigma}$ error bars from][]{tr18} compared to the surface evolution of Set 0 and Set 2 stellar models of solar (top panel) and of double-solar metallicity (bottom panel) of different masses (see legend). Circles (Set 0) and triangles (Set 2) on the lines represent the composition after each TDU after the envelope becomes C-rich. The black dashed lines represent the solar composition ($\mathrm{\delta}$ = 0 by definition).}
      \label{figure:figure 3}
	\end{center}
\end{figure} 

\subsubsection{\texorpdfstring{$\mathrm{^{64}Ni}$}{}}
\label{section:section 3.3.1.}

In the case of $\mathrm{^{64}Ni}$, the ratio between its ASTRAL and KaDoNiS neutron-capture cross sections decreases with temperature (see Table \ref{table:6}). In general, the neutron-capture rate based on ASTRAL is the highest of the two during the $\mathrm{^{13}C(\alpha,n)}$ phases, while the rate based on the MACS from KaDoNiS is the higher during the $\mathrm{^{22}Ne(\alpha,n)}$ phases. In most cases, this results in a 6\%-20\% reduction in the abundance of $\mathrm{^{64}Ni}$ in the Set 2 models compared to the Set 1 models, except for the 4 $\mathrm{M_{\odot}}$ \textit{Z} = 0.007 model, where there was no difference between the predictions.

\begin{figure*} [ht!]
	\begin{center}
		\includegraphics[width=\hsize]{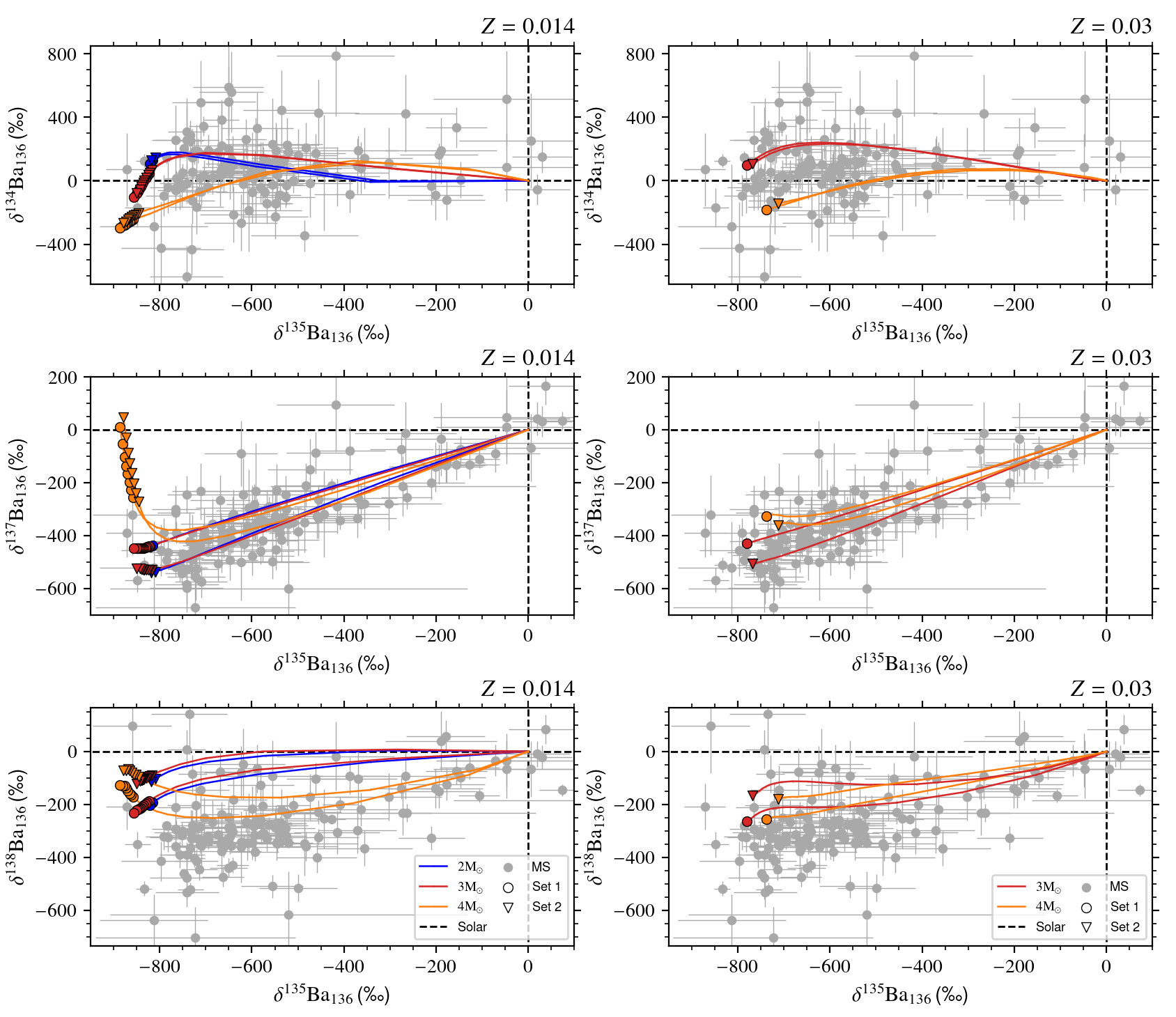}
		\caption{$\mathrm{\delta^{134}Ba_{136}}$ (top row), $\mathrm{\delta^{137}Ba_{136}}$ (middle row) and $\mathrm{\delta^{138}Ba_{136}}$ (bottom row) versus $\mathrm{\delta^{135}Ba_{136}}$ from mainstream SiC grain data \citep[dark gray circles with 1$\mathrm{\sigma}$ error bars from][]{sa03, ba07, ma07, li14, li15, li17b, li17, li19, li22, st18, st19} compared to the surface evolution of Set 1 and Set 2 stellar models of solar (left panels) and of double-solar metallicity (right panels) of different masses (see legend). Circles (Set 1) and triangles (Set 2) on the lines represent the composition after each TDU after the envelope becomes C-rich. The black dashed lines represent the solar composition ($\mathrm{\delta}$ = 0 by definition).}
      \label{figure:figure 4}
	\end{center}
\end{figure*}

Of the five stable Ni isotopes, the most abundant $\mathrm{^{58}Ni}$ (which was used as reference isotope for the stardust grain isotopic ratio data) and $\mathrm{^{60}Ni}$ are not affected by \textit{s}-process nucleosynthesis, in contrast to the much more rare $\mathrm{^{61,62,64}Ni}$ isotopes (1.14\%, 3.63\% and 0.93\% of total Ni abundance, respectively), which can be produced by neutron captures. In Fig. \ref{figure:figure 3}, we compare $\mathrm{\delta^{60}Ni_{58}}$ versus $\mathrm{\delta^{64}Ni_{58}}$ from \citet{tr18} to the predictions of the Set 0 and Set 2 models, since the final surface abundance of $\mathrm{^{64}Ni}$ also changes between Set 1 and Set 0 (see Sect. \ref{section:section 3.1.1.}). We see a relatively large reduction (168-724\textperthousand, depending on the stellar model and thermal pulse) in $\mathrm{\delta^{64}Ni_{58}}$ from Set 2 to Set 0 during the stellar evolution. The Set 0 models (except for 2 $\mathrm{M_{\odot}}$ \textit{Z} = 0.014 model) overproduce $\mathrm{^{64}Ni}$ relative to grain data at the end of the evolution, while Set 2 models can better cover grain data. Note that the $\mathrm{\delta^{60}Ni_{58}}$ values do not change between Set 2 and Set 0. Their fluctuation in the stardust grain data is probably due to GCE as the $\mathrm{\delta^{60}Ni_{58}}$ values do not change significantly in AGB models \citep{tr18}. 

\subsubsection{\texorpdfstring{$\mathrm{^{134,135,136,137,138}Ba}$}{}}
\label{section:section 3.3.2.}

Among the five barium isotopes on the main \textit{s}-process path, the final surface abundance of $\mathrm{^{134,136,137}Ba}$ decreases significantly in the Set 2 models compared to the Set 1 models, consistent with the change in reaction rates (see Table \ref{table:6}). The only exception is the 4 $\mathrm{M_{\odot}}$ \textit{Z} = 0.007 model, where the final surface abundance of $\mathrm{^{137}Ba}$ slightly increases from Set 1 to Set 2, probably due to some effects related to the $\mathrm{^{137}Cs}$ branching point. 

When we compare the mainstream SiC grain data to our predictions from the Set 1 and Set 2 models, we conclude the following: (1) The surface abundances of the \textit{s}-only $\mathrm{^{136}Ba}$ used as a reference isotope (as commonly used in stardust grain analysis) are lower in Set 2 models compared to Set 1 models. Following this, the $\mathrm{\delta^{135}Ba_{136}}$ values slightly increase (by 5-11\textperthousand), except for the 4 $\mathrm{M_{\odot}}$ \textit{Z} = 0.03 model, where it increases more significantly (by 26\textperthousand). For the same reason, the $\mathrm{\delta^{134}Ba_{136}}$ and $\mathrm{\delta^{138}Ba_{136}}$ values also increase (by 7-38\textperthousand\, and 55-110\textperthousand, respectively), despite the fact that the surface abundances of $\mathrm{^{134}Ba}$ also decrease, and the neutron-capture rate of $\mathrm{^{138}Ba}$ does not change. (2) The $\mathrm{\delta^{137}Ba_{136}}$ values decrease by 76-100\textperthousand\, (and by -32\textperthousand\, in 4 $\mathrm{M_{\odot}}$ \textit{Z} = 0.03 model) because the cross section of $\mathrm{^{137}Ba}$ increase more than the cross section of $\mathrm{^{136}Ba}$ (see Table \ref{table:6}). The exception is the 4 $\mathrm{M_{\odot}}$ \textit{Z} = 0.014 model, where $\mathrm{^{137}Ba}$ is overproduced (also relative to stardust grain data) due to activation of branching points at $\mathrm{^{134,135,136}Cs}$ \citep[more details in][]{lu18a}. \citet{ve21} compared the prediction of the original to the ASTRAL-based neutron-capture data set of the FRUITY Magnetic models \citep{ve20}, and found similar results to ours in the cases of $\mathrm{\delta^{137}Ba_{136}}$ and $\mathrm{\delta^{138}Ba_{136}}$.

\begin{figure} [ht!]
	\begin{center}
		\includegraphics[width=\hsize]{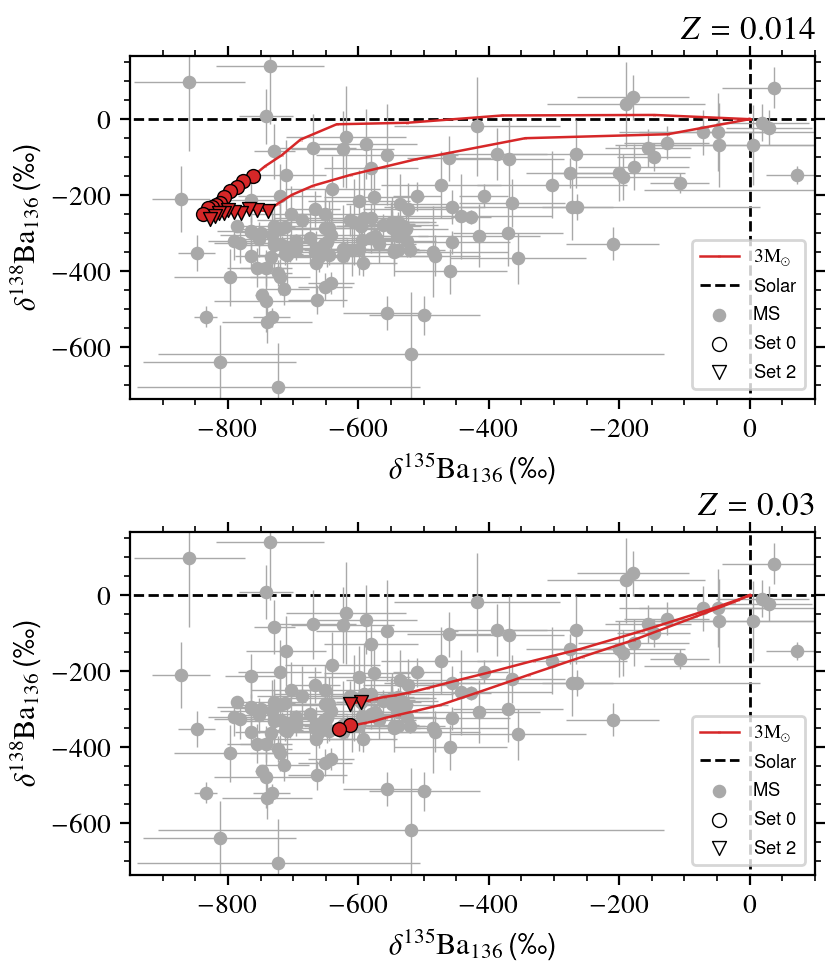}
		\caption{Same as the bottom row of Fig. \ref{figure:figure 4}, but including only \textit{M} = 3 M$_{\odot}$ models calculated with $M_{\mathrm{PMZ}}$ = 5 $\times$ $10^{-4}$ M$_{\odot}$}.
      \label{figure:figure 7}
	\end{center}
\end{figure}

Comparing our predictions to the grain data, the conclusions of \citet[][based on the predictions of the Set 0 models]{lu18a} have not changed significantly. The two main differences are: (1) Set 2 models provide us with a better fit to the bulk of the $\mathrm{\delta^{137}Ba_{136}}$ versus $\mathrm{\delta^{135}Ba_{136}}$ stardust grain data. (2) However, the \textit{Z} = 0.03 models of Set 2 produce higher $\mathrm{\delta^{138}Ba_{136}}$ than the typical values of SiC grains, which was previously only true for solar metallicity models. According to \citet{lu18a}, the $\mathrm{\delta^{138}Ba_{136}}$ ratios can be reduced by decreasing the $M_{\mathrm{PMZ}}$ parameter. Calculating the \textit{M} = 3 $\mathrm{M_{\odot}}$ models with a lower $M_{\mathrm{PMZ}}$ than the "standard" value ($M_{\mathrm{PMZ}}$ = 5 $\times$ $10^{-4}$ M$_{\odot}$ instead of 2 $\times$ $10^{-3}$ M$_{\odot}$), the predictions of 3 $\mathrm{M_{\odot}}$ \textit{Z} = 0.03 model become more consistent with the grain data in both Set 1 and Set 2 (Fig. \ref{figure:figure 7}, bottom panel), although, as discussed in \citet{lu18a}, more efficient TDU of material with such composition is needed to reach the most extreme data points. However, the 3 $\mathrm{M_{\odot}}$ \textit{Z} = 0.014 model still produces higher $\mathrm{\delta^{138}Ba_{136}}$ ratios than the typical values of SiC grains (Fig. \ref{figure:figure 7}, top panel).

\subsubsection{\texorpdfstring{$\mathrm{^{186}W}$}{}}
\label{section:section 3.3.3.}

\begin{figure} [ht!]
	\begin{center}
		\includegraphics[width=\hsize]{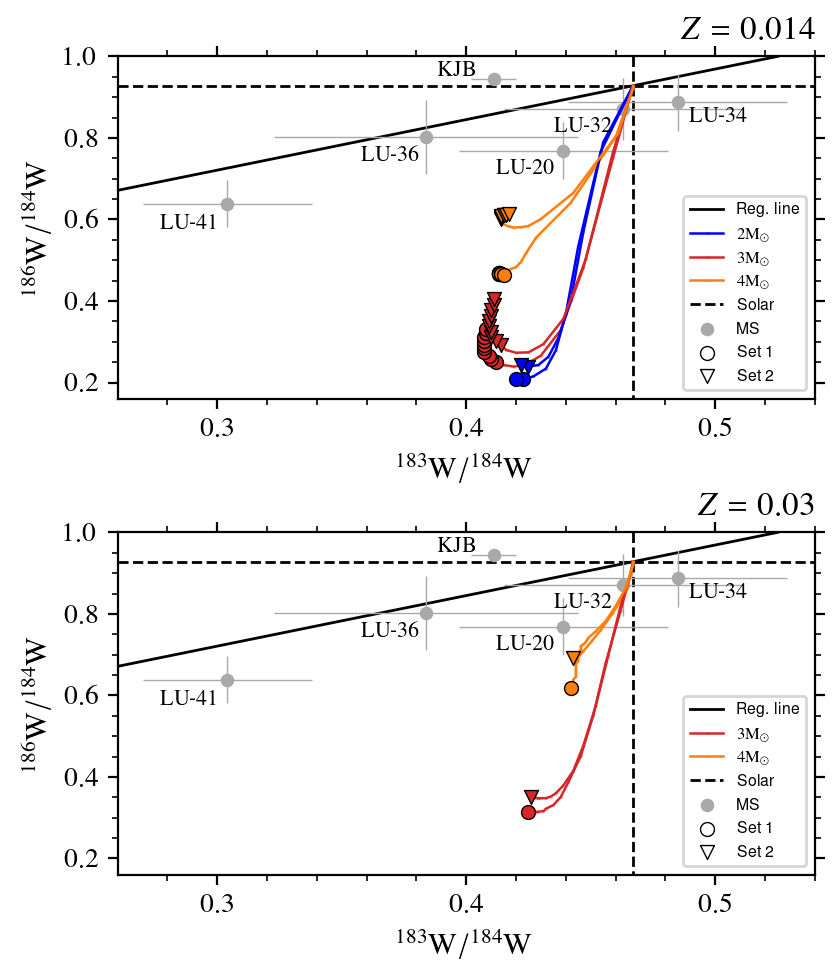}
		\caption{$\mathrm{^{186}W/^{184}W}$ versus $\mathrm{^{183}W/^{184}W}$ from single large grains (LU) and the average of small grains \citep[KJB, dark gray circles with 1$\mathrm{\sigma}$ error bars from][]{av12} compared to the surface evolution of Set 1 and Set 2 stellar models of solar (top panel) and of double-solar metallicity (bottom panel) of different masses (see legend). Circles (Set 0) and triangles (Set 2) on the lines represent the composition after each TDU after the envelope becomes C-rich. The black dashed lines represent the solar composition ($\mathrm{\delta}$ = 0 by definition). The solid black line is the regression line from \citet{av12}.}
      \label{figure:figure 5}
	\end{center}
\end{figure}

Due to the ASTRAL-based neutron-capture rate of $\mathrm{^{186}W}$ being lower than the JINA reaclib fit to the KaDoNiS data, the Set 2 models obtain 1.07-1.44 times more $\mathrm{^{186}W}$ than the Set 1 models at the end of the evolution. \citet{av12} reported the first tungsten isotopic measurements in SiC grains and compared the isotopic ratios to the predictions of FRUITY and the then version of the \textit{Monash} models\footnote{Stellar models: \textit{M} = 1.25, 1.8 and 3 $\mathrm{M_{\odot}}$ \textit{Z} = 0.01 and \textit{M} = 3 and 4 $\mathrm{M_{\odot}}$ \textit{Z} = 0.02.}. They found that most models predict lower $\mathrm{^{186}W}$/$\mathrm{^{184}W}$ ratios than those measured in the grains. To check whether our new model predictions change this conclusion, we plot the $\mathrm{^{186}W}$/$\mathrm{^{184}W}$ versus $\mathrm{^{183}W}$/$\mathrm{^{184}W}$ predictions and data in Fig. \ref{figure:figure 5}. The Set 2 models provided us with higher $\mathrm{^{186}W}$/$\mathrm{^{184}W}$ ratios than the Set 1 models, in addition, the 4 $\mathrm{M_{\odot}}$ \textit{Z} = 0.014 model resulted in a similar $\mathrm{^{186}W}$ /$\mathrm{^{184}W}$ ratios as the LU-41 grain. However, no models are able to match full composition of LU-41, nor the very high $\mathrm{^{186}W}$/$\mathrm{^{184}W}$ ratio shown by the bulk KJB data. It should be noted, however, that this result strongly depends on the neutron-capture rate of the branching point isotope $\mathrm{^{185}W}$, for which no direct measurements are available, and current estimates are derived from inverse photodisintegration experiments \citep{so03,mo04}. 

Further constrains on the \textit{s}-process isotopic composition of W come from other types of meteoritic materials. From analysis of chondrules and matrix of the Allende meteorite, \citet{budde16} obtained a slope of +1.25 $\pm$ 0.06 (2$\sigma$) for the relationship between $\mathrm{^{182}W}$/$\mathrm{^{184}W}$ and $\mathrm{^{183}W}$/$\mathrm{^{184}W}$ \citep[when internally normalised to $\mathrm{^{186}W}$/$\mathrm{^{184}W}$ and the solar ratios, see, e.g.][for full details]{lugaro23}, and \citet{kruijer14} obtained +1.41 $\pm$ 0.05 from analysis of bulk calcium-aluminium-rich inclusions (CAIs). In Fig. \ref{figure:plotW}, we compare the highest and lowest observed slope, derived when considering the error bars, to the slopes derived by selected Set 2 models. The temperature conditions required to match the observations are found for stellar models in range between the 3 M$_{\odot}$ \textit{Z} = 0.014 and the 4 M$_{\odot}$ \textit{Z} = 0.03 models. The results do not change significantly when considering Set 1 models. Instead, they change significantly depending on the addition of the abundance of radioactive $\mathrm{^{182}Hf}$ to the abundance of $\mathrm{^{182}W}$. In fact, the observed slopes can be matched only if $\mathrm{^{182}Hf}$ is fully included (as done in Fig. \ref{figure:plotW}), otherwise all the lines have slopes between 1.8 and 2.4, much higher than observed. This indicated that Hf must have fully condensed together with W in the stardust grains that carried these anomalies into the early Solar System. Another interesting result is that when we force, for example, the 3 M$_{\odot}$ \textit{Z} = 0.014 model to match the stardust data shown in Fig. \ref{figure:figure 5}, by decreasing the abundance of $\mathrm{^{183}W}$ by 0.7-0.85 and/or increasing the abundance of $\mathrm{^{186}W}$ by a factor of 2, the slope always decrease to values outside the observed range (because \textit{s}-process addition produces negative values). 

This indicates that the stardust grains that carried the anomalies represented in Fig. \ref{figure:plotW} are not the same stardust grains plotted in Fig. \ref{figure:figure 5}. This may not be surprising given that the plotted single LU grains have very large size (roughly 10-20 $\mu$m), and are probably be quite rare, while the bulk grain KJB data point represents a grand average of grains of small size ($\sim$ 0.5 $\mu$m). Instead, the chondrules, matrix, and CAIs data indicate a carrier with the signature of the nucleosynthesis produced in AGB models of mass and metallicity similar to those that match the composition of the single KJG grains, of typical $\mu$m size \citep[e.g.][]{lu18a}. It is also possible that other type of grains than SiC carried the anomalies shown, as already suggested for example by \citet{schonbachler05,reisberg09} on the basis of leachates experiments. 

\section{Conclusions}
\label{section:section 4}
We studied how temperature- and density-dependent (instead of constant, terrestrial) decay rates and new neutron-capture cross sections affect \textit{s}-process nucleosynthesis predictions. We provide the first database of surface abundances and stellar yields of the isotopes heavier than iron from the \textit{Monash} models. We used seven stellar structure models \citep{ka14,ka16} of low-mass AGB stars, and three sets of the nuclear network (Sets 0, 1, and 2) based on three databases: JINA reaclib \citep{cy10}, NETGEN \citep{jo01,ai05,xu13}, and ASTRAL \citep{re18,ve21,ve23}. We compared the final surface abundances of the isotopes for each set and discussed the relevant cases. We calculated the solar \textit{s}-process contribution for the \textit{p}-only isotopes $\mathrm{^{94}Mo}$, $\mathrm{^{108}Cd}$ and $\mathrm{^{152}Gd}$, as well as for the \textit{s}-only isotopes $\mathrm{^{176}Lu}$ and $\mathrm{^{176}Hf}$, the radioactive-to-stable abundance ratios for the short-lived radionuclides $\mathrm{^{129}I}$, $\mathrm{^{182}Hf}$ and $\mathrm{^{205}Pb}$, and the isolation time of the Solar System using a steady-state equation and the $\mathrm{^{205}Pb}/\mathrm{^{204}Pb}$ ratio. We compared our solar metallicity models with the same FRUITY \citep{cr09,cr11,cr15} models, which also uses \textit{T}- and $\rho$-dependent decay rates. In addition to the above, we compared the predictions of the solar and double-solar metallicity models with mainstream presolar stardust SiC grain data for the $\mathrm{^{60}Ni}$/$\mathrm{^{64}Ni}$, $\mathrm{^{80}Kr}$/$\mathrm{^{82}Kr}$, $\mathrm{^{94}Mo}$/$\mathrm{^{96}Mo}$, $\mathrm{^{134,135,137,138}Ba}$/$\mathrm{^{136}Ba}$,
and $\mathrm{^{186}W}$/$\mathrm{^{184}W}$ ratios. 

\begin{figure} [ht!]
	\begin{center}
		\includegraphics[width=\hsize]{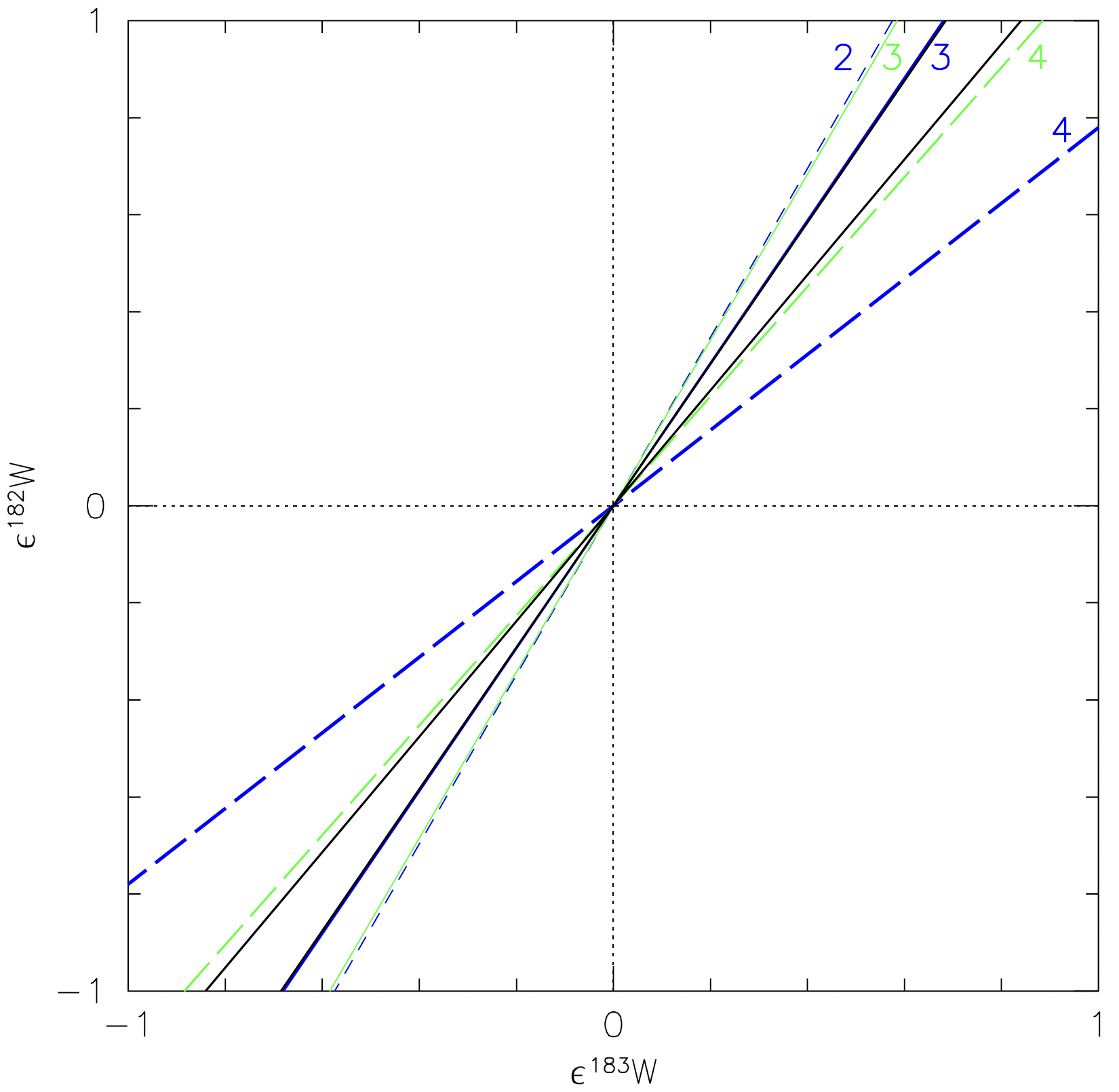}
		\caption{Linear correlations between $\epsilon\mathrm{^{182}W}$ (including the $\mathrm{^{182}Hf}$ abundance) and $\epsilon\mathrm{^{183}W}$ (i.e. ratios internally normalised to $\mathrm{^{186}W}$/$\mathrm{^{184}W}$ and relative to the solar ratios, multiplied by 10,000) derived from the lowest and the upper limit of the observations (solid black lines) and from the abundances calculated at the end of the evolution of Set 2 models of \textit{Z} = 0.014 (blue lines) and \textit{Z} = 0.03 (green lines), with stellar mass indicated by corresponding labels and and different line types: 2 M$_{\odot}$ (short-dashed), 3 M$_{\odot}$ (solid), and 4 M$_{\odot}$ (long-dashed). Note that the \textit{M} = 3 M$_{\odot}$ \textit{Z} = 0.014 line overlaps with the line of the observational upper limit. The dotted lines indicate the solar composition. Addition of \textit{s}-process material produces negative values, while removal correspond to positive values.}
      \label{figure:plotW}
	\end{center}
\end{figure} 
Our main conclusions are: 
\begin{enumerate}
    \item[(1)] Taking into account the \textit{T} and $\rho$ dependence of the decay rates of the $\mathrm{^{79}Se}$ and $\mathrm{^{80}Br}$ branching isotopes (Set 1), our predictions become more consistent with the intershell $\mathrm{^{80}Kr}$/$\mathrm{^{82}Kr}$ ratios derived by \citet{le94} from stardust SiC grain data than the Set 0 models (see Table \ref{table:4}). 
    \item[(2)] Taking into account the \textit{T} and $\rho$ dependence of the decay rates of the $\mathrm{^{93}Zr}$ and $\mathrm{^{94}Nb}$ branching isotopes (Set 1), the surface evolution of the 2 and 3 $\mathrm{M_{\odot}}$ solar and double-solar metallicity models follow closely the regression line of the $[\mathrm{^{94}Mo/^{96}Mo}]$ versus $[\mathrm{^{92}Mo/^{96}Mo}]$ SiC grain data (see Fig. \ref{figure:figure 1}). Our models quantitatively reproduce the minor $s$-process production of the classical p-only $\mathrm{^{94}Mo}$ observed in stardust SiC grains, of around 3-4\%.
    \item[(3)] Taking into account the \textit{T} and $\rho$ dependence of the decay rates of the $\mathrm{^{151}Sm}$ and $\mathrm{^{152}Eu}$ branching isotopes (Set 1), we predict a high, 0.7 solar \textit{s}-process contribution for the traditionally \textit{p}-only $\mathrm{^{152}Gd}$, which is consistent with results from previous studies \citep{ar99,bi11,bi14}.
    \item[(4)] Our Set 1 models do not reproduce the Solar System ratio of the two \textit{s}-only isotopes $\mathrm{^{176}Lu}$ and $\mathrm{^{176}Hf}$ due to the NETGEN rate of $\mathrm{^{176}Lu}$ incorrectly assuming thermal equilibrium between the long-lived ground state and the short-lived isomer over the whole temperature range. This problem can be solved by using a different \textit{T}-dependent $\mathrm{\beta}$-decay rate for $\mathrm{^{176}Lu}$ \citep[e.g. from][]{he08a}, but note that the coupling between the ground state and the isomer is still uncertain and needs to be revised.
    \item[(5)] Taking into account the \textit{T} and $\rho$ dependence of the decay rates of $\mathrm{^{205}Tl}$, $\mathrm{^{204}Pb}$ and $\mathrm{^{205}Pb}$ (Set 1), we cannot obtain a positive solution for the isolation time of the Solar System using the $\mathrm{^{205}Pb}/\mathrm{^{204}Pb}$ ratio from solar metallicity models, suggesting that the rates of the $\mathrm{^{205}Tl}$-$\mathrm{^{205}Pb}$ decay system need to be revised. This was recently done by \citet{le24}.
    \item[(6)] With the new (n,$\gamma$) cross sections of $\mathrm{^{64}Ni}$ and \textit{T}- and $\rho$-dependent decay rates of $\mathrm{^{63}Ni}$ and $\mathrm{^{64}Cu}$ (Set 2 models), the $\mathrm{\delta^{64}Ni_{58}}$ values derived from the solar and double-solar metallicity models are smaller than those of Set 0, and better cover the SiC grain data (see Fig. \ref{figure:figure 3}).
    \item[(7)] With the new (n,$\gamma$) cross sections of $\mathrm{^{136,137}Ba}$ (Set 2), the solar and double-solar metallicity models provide us with a better fit to the $\mathrm{\delta^{137}Ba_{136}}$ SiC grain data, but produce higher $\mathrm{\delta^{138}Ba_{136}}$ than the typical values of the grains than those of the previous sets (see Fig. \ref{figure:figure 4}).
    \item[(8)] The Set 2 models produce higher $\mathrm{^{186}W}$, although they are still not matching well the W isotopic composition of large SiC stardust grains (Fig. \ref{figure:figure 5}). However, they do match well the W composition observed in other types of meteoritic materials (Fig. \ref{figure:plotW}).
\end{enumerate}

Future plans involve extending our work based on recent and future measurements of $\mathrm{\beta}$-decay rates and (n,$\gamma$) cross sections. The PANDORA project \citep[Plasmas for Astrophysics, Nuclear Decays Observation and Radiation for Archaeometry,][]{ma22} will be a possible source of $\mathrm{\beta}$-decay rates from plasma measurements. The first experimental campaign involves three important isotopes for the \textit{s} process: $\mathrm{^{94}Nb}$, $\mathrm{^{134}Cs}$, and $\mathrm{^{176}Lu}$. The CERN neutron time-of-flight facility, n\_TOF \citep{gu13} has been providing neutron-capture cross sections for various isotopes for many years. The most recent publications present the (n,$\gamma$) cross-section measurements of $\mathrm{^{89}Y}$ \citep{ta24}, $\mathrm{^{94}Nb}$ \citep{ba23}, $\mathrm{^{140}Ce}$ \citep{am24}, and $\mathrm{^{204}Tl}$ \citep{ca24} that are relevant for the \textit{s} process. The effect of the $\mathrm{^{22}Ne(\alpha,n)^{25}Mg}$ reaction rate and its uncertainties well need to be tested. Direct cross-section measurement of this neutron-source reaction is currently been carried out at the LUNA \citep[Laboratory for Underground Nuclear Astrophysics,][]{an24}. In addition, we plan to calculate more models with different masses and metallicities to be used as input for the galactic chemical evolution of the isotopes of the elements heavier than iron and to carry out further detailed comparison with meteoritic data.

\section{Data availability}
Surface isotopic abundance and stellar yield data tables of the seven stellar models of low-mass AGB stars for the three nuclear networks are available on Zenodo through the following link: \href{https://zenodo.org/records/14981333}{https://zenodo.org/records/14981333}.

\begin{acknowledgements}
This project has been supported by the Lend\"ulet Program LP2023-10 of the Hungarian Academy of Sciences. This work was supported by the NKFIH Excellence Grant TKP2021-NKTA-64. Amanda Karakas was supported by the Australian Research Council Centre of Excellence for All Sky Astrophysics in 3 Dimensions (ASTRO 3D), through project number CE170100013. Balázs Szányi thanks ASTRO 3D for supporting his travel to Monash University. The authors thank Marco Pignatari for his constructive comments throughout this project. The authors thank the anonymous referee for their careful comments, which helped to improve the paper.
\end{acknowledgements}
\bibliographystyle{aa}
\bibliography{first}

\begin{appendix}
\onecolumn
\section{Additional tables}
\setlength{\tabcolsep}{1em}
\begin{table*}[h!]
\caption{$Y^\mathrm{i}_{\mathrm{Set\,1,final}}$/$Y^\mathrm{i}_{\mathrm{Set\,0,final}}$ for the 30 isotopes that vary the most between the two sets.}
\label{table:A.1.}
\centering
\begin{tabular}{cccccccc}
\hline
Isotope & \multicolumn{7}{c}{Stellar model} \\
& \multicolumn{2}{c}{\textit{Z} = 0.007} & \multicolumn{3}{c}{\textit{Z} = 0.014} & \multicolumn{2}{c}{\textit{Z} = 0.03} \\\cline{2-8}
& 3\,$\mathrm{M_{\odot}}$ & 4\,$\mathrm{M_{\odot}}$ & 2\,$\mathrm{M_{\odot}}$ & 3\,$\mathrm{M_{\odot}}$ & 4\,$\mathrm{M_{\odot}}$ & 3\,$\mathrm{M_{\odot}}$ & 4\,$\mathrm{M_{\odot}}$ \\
\hline
$\mathrm{^{64}Ni}$	&	\textbf{0.88}	&	0.99	&	0.97	&	\textbf{0.89}	&	0.96	&	\textbf{0.89}	&	\textbf{0.90}	\\
$\mathrm{^{71}Ga}$	&	\textbf{0.89}	&	\textbf{0.83}	&	0.98	&	0.93	&	\textbf{0.87}	&	0.98	&	\textbf{0.90}	\\
$\mathit{^{79}Se}$	&	\textit{\textbf{0.83}}	&	\textit{\textbf{0.82}}	&	\textit{\textbf{0.56}}	&	\textit{\textbf{0.77}}	&	\textit{\textbf{0.80}}	&	\textit{\textbf{0.64}}	&	\textit{\textbf{0.78}}	\\
$\mathrm{^{80}Se}$	&	\textbf{0.90}	&	1.01	&	0.91	&	\textbf{0.88}	&	0.97	&	\textbf{0.89}	&	0.93	\\
$\mathrm{^{79}Br}$	&	1.03	&	\textbf{1.23}	&	0.99	&	0.99	&	\textbf{1.15}	&	1.00	&	\textbf{1.13}	\\
$\mathrm{^{80}Kr}$	&	\textbf{2.65}	&	\textbf{1.94}	&	\textbf{2.69}	&	\textbf{3.81}	&	\textbf{1.87}	&	\textbf{4.62}	&	\textbf{2.31}	\\
$\mathit{^{81}Kr}$	&	\textit{\textbf{13.00}}	&	\textit{\textbf{355.43}}	&	\textit{\textbf{8.51}}	&	\textit{\textbf{11.24}}	&	\textit{\textbf{87.88}}	&	\textit{\textbf{12.46}}	&	\textit{\textbf{31.80}}	\\
$\mathrm{^{82}Kr}$	&	1.06	&	1.04	&	1.08	&	\textbf{1.12}	&	1.02	&	\textbf{1.15}	&	1.09	\\
$\mathit{^{94}Nb}$	&	\textit{\textbf{0.03}}	&	\textit{\textbf{0.03}}	&	\textit{\textbf{0.01}}	&	\textit{\textbf{0.01}}	&	\textit{\textbf{0.03}}	&	\textit{\textbf{0.01}}	&	\textit{\textbf{0.02}}	\\
$\mathrm{^{94}Mo}$	&	\textbf{1.51}	&	1.01	&	\textbf{1.33}	&	\textbf{1.53}	&	1.02	&	\textbf{1.23}	&	1.04	\\
$\mathrm{^{108}Cd}$	&	\textbf{1.32}	&	1.02	&	\textbf{1.62}	&	\textbf{1.49}	&	1.02	&	\textbf{1.30}	&	1.03	\\
$\mathit{^{129}I}$	&	\textit{\textbf{0.20}}	&	\textit{\textbf{0.49}}	&	\textit{\textbf{0.20}}	&	\textit{\textbf{0.19}}	&	\textit{\textbf{0.33}}	&	\textit{\textbf{0.16}}	&	\textit{\textbf{0.23}}	\\
$\mathrm{^{134}Xe}$	&	0.95	&	\textbf{0.89}	&	\textbf{1.27}	&	\textbf{1.18}	&	\textbf{0.87}	&	\textbf{1.10}	&	0.98	\\
$\mathrm{^{148}Nd}$	&	0.93	&	\textbf{0.89}	&	0.96	&	0.91	&	0.92	&	0.97	&	0.96	\\
$\mathrm{^{152}Gd}$	&	\textbf{24.79}	&	\textbf{27.56}	&	\textbf{7.41}	&	\textbf{11.32}	&	\textbf{9.17}	&	\textbf{4.79}	&	\textbf{2.18}	\\
$\mathrm{^{170}Er}$	&	\textbf{0.87}	&	\textbf{0.81}	&	0.93	&	\textbf{0.86}	&	\textbf{0.86}	&	0.96	&	0.94	\\
$\mathrm{^{171}Yb}$	&	\textbf{0.89}	&	\textbf{0.81}	&	0.95	&	\textbf{0.90}	&	\textbf{0.85}	&	0.95	&	0.93	\\
$\mathit{^{176}Lu}$	&	\textit{\textbf{0.51}}	&	\textit{\textbf{0.50}}	&	\textit{\textbf{0.54}}	&	\textit{\textbf{0.49}}	&	\textit{\textbf{0.49}}	&	\textit{\textbf{0.45}}	&	\textit{\textbf{0.50}}	\\
$\mathrm{^{176}Hf}$	&	\textbf{45.68}	&	\textbf{41.56}	&	\textbf{10.25}	&	\textbf{15.09}	&	\textbf{8.94}	&	\textbf{4.30}	&	\textbf{1.85}	\\
$\mathit{^{182}Hf}$	&	0.97	&	0.95	&	\textbf{0.79}	&	\textbf{0.89}	&	0.98	&	\textbf{0.82}	&	0.91	\\
$\mathrm{^{186}W}$	&	0.93	&	\textbf{0.86}	&	\textbf{0.89}	&	\textbf{0.90}	&	\textbf{0.90}	&	0.93	&	0.94	\\
$\mathrm{^{187}Os}$	&	\textbf{1.13}	&	\textbf{1.12}	&	1.03	&	1.07	&	\textbf{1.13}	&	1.03	&	1.03	\\
$\mathrm{^{192}Os}$	&	\textbf{0.73}	&	\textbf{0.65}	&	0.94	&	\textbf{0.85}	&	\textbf{0.87}	&	0.97	&	0.98	\\
$\mathrm{^{192}Pt}$	&	\textbf{1.11}	&	\textbf{1.07}	&	1.05	&	1.07	&	\textbf{1.10}	&	1.04	&	1.03	\\
$\mathit{^{205}Tl}$	&	\textit{\textbf{2.42}}	&	\textit{\textbf{0.62}}	&	\textit{\textbf{7.33}}	&	\textit{\textbf{4.27}}	&	\textit{\textbf{0.90}}	&	\textit{\textbf{3.64}}	&	\textit{\textbf{1.26}}	\\
$\mathrm{^{204}Pb}$	&	\textbf{1.20}	&	\textbf{1.58}	&	1.02	&	1.08	&	\textbf{1.43}	&	1.02	&	1.08	\\
$\mathit{^{205}Pb}$	&	\textit{\textbf{0.07}}	&	\textit{\textbf{3.60}}	&	\textit{\textbf{0.03}}	&	\textit{\textbf{0.04}}	&	\textit{\textbf{0.77}}	&	\textit{\textbf{0.01}}	&	\textit{\textbf{0.05}}	\\
$\mathrm{^{208}Pb}$	&	0.95	&	\textbf{0.82}	&	0.94	&	0.93	&	0.97	&	0.95	&	0.97	\\
$\mathrm{^{209}Bi}$	&	0.96	&	\textbf{0.77}	&	0.96	&	0.95	&	0.97	&	0.99	&	1.00	\\
$\mathrm{^{210}Po}$	&	0.94	&	\textbf{0.63}	&	0.91	&	0.91	&	0.94	&	\textbf{0.89}	&	\textbf{0.88}	\\
\hline
\end{tabular}
\tablefoot{$Y^\mathrm{i}_{\mathrm{Set\,1,final}}$/$Y^\mathrm{i}_{\mathrm{Set\,2,final}}$ indicates the ratio of the final surface normalised number abundance of isotope \textit{i} between Set 1 and Set 0. We list only those isotopes that abundance changed at least 10\% (in bold) in at least one Set 1 model compared to the same Set 0 model. The unstable isotopes are in italics.}
\end{table*}
\setlength{\tabcolsep}{0.32em}
\begin{table*}
\caption{Relevant information of each isotope case discussed in Sect. \ref{section:section 3.1.}.} 
\label{table:table 3} 
\centering
\begin{tabular}{cccccccccc}
\hline
\hline
Isotope & \makecell{Branching \\ isotope} & \makecell{Decay \\ mode} & \makecell{Daughter \\ nucleus} & \makecell{$\lambda_{\mathrm{Set\,0}}$ \\($\mathrm{s^{-1}}$)} & \makecell{$\mathrm{n_e}$ \\ ($\mathrm{10^{26}}$ $\mathrm{cm^{-3}}$)} & \makecell{$\lambda^{\mathrm{90\,MK}}_{\mathrm{Set\,1}}$ \\($\mathrm{s^{-1}}$)} & \makecell{$\lambda^{\mathrm{300\,MK}}_{\mathrm{Set\,1}}$ \\($\mathrm{s^{-1}}$)} & \makecell{Direction \\ of change} & Relevance \\
\hline
$\mathrm{^{64}Ni}$ & & & & stable & & stable & stable & $\downarrow$ & Compare to stardust \\
& $\mathrm{^{63}Ni}$ & $\mathrm{\beta^-}$ & $\mathrm{^{63}Cu}$ & 2.2 $\times$ $\mathrm{10^{-10}}$ & 1 & 5.2 $\times$ $\mathrm{10^{-10}}$ & 1.8 $\times$ $\mathrm{10^{-9}}$ & & SiC grain data \\
& & & & & 9.5 & 3.8 $\times$ $\mathrm{10^{-10}}$ & 1.7 $\times$ $\mathrm{10^{-9}}$ & & \\
& $\mathrm{^{64}Cu}$ & $\mathrm{\beta^+}$ & $\mathrm{^{64}Ni}$ & 9.2 $\times$ $\mathrm{10^{-6}}$ & 1 & 3.2 $\times$ $\mathrm{10^{-6}}$ & 2.7 $\times$ $\mathrm{10^{-6}}$ & & \\
& & & & & 9.5 & 5.8 $\times$ $\mathrm{10^{-6}}$ & 3.2 $\times$ $\mathrm{10^{-6}}$ & & \\
& & $\mathrm{\beta^-}$ & $\mathrm{^{64}Zn}$ & 5.9 $\times$ $\mathrm{10^{-6}}$ & 1 & 6.3 $\times$ $\mathrm{10^{-6}}$ & 6.4 $\times$ $\mathrm{10^{-6}}$ & & \\
& & & & & 9.5 & 6.1 $\times$ $\mathrm{10^{-6}}$ & 6.3 $\times$ $\mathrm{10^{-6}}$ & & \\
\hline
$\mathrm{^{80}Kr}$ & & & & stable & & stable & stable & $\uparrow$ & Compare to stardust \\
$\mathrm{^{81}Kr}$ & & $\mathrm{\epsilon}$& $\mathrm{^{81}Br}$ & 9.6 $\times$ $10^{-14}$ & 1 & 1.6 $\times$ $10^{-14}$ & 1.6 $\times$ $10^{-12}$ & $\uparrow$ & SiC grain data \\
& & & & & 9.5 & 6.4 $\times$ $10^{-14}$ & 4.1 $\times$ $10^{-11}$ & & \\
& $\mathrm{^{79}Se}$ & $\mathrm{\beta^-}$ & $\mathrm{^{79}Br}$ & 7.4 $\times$ $\mathrm{10^{-14}}$ & 1 & 2.7 $\times$ $\mathrm{10^{-12}}$ & 1.3 $\times$ $\mathrm{10^{-8}}$ & & \\
& & & & & 9.5 & 2.3 $\times$ $\mathrm{10^{-12}}$ & 1.3 $\times$ $\mathrm{10^{-8}}$ & & \\
& $\mathrm{^{80}Br}$ & $\mathrm{\beta^-}$ & $\mathrm{^{80}Kr}$ & 6.0 $\times$ $\mathrm{10^{-4}}$ & 1 & 5.9 $\times$ $\mathrm{10^{-4}}$ & 3.9 $\times$ $\mathrm{10^{-4}}$ & & \\
& & $\mathrm{\beta^+}$ & $\mathrm{^{80}Se}$ & 5.4 $\times$ $\mathrm{10^{-5}}$ & 1 & 2.5 $\times$ $\mathrm{10^{-5}}$ & 1.1 $\times$ $\mathrm{10^{-5}}$ & & \\
& & & & & 9.5 & 4.7 $\times$ $\mathrm{10^{-5}}$ & 1.3 $\times$ $\mathrm{10^{-5}}$ & & \\
\hline
$\mathrm{^{94}Mo}$ & & & & stable & & stable & stable & $\uparrow$ & \textit{p}-only \\
& $\mathrm{^{93}Zr}$ & $\mathrm{\beta^-}$ & $\mathrm{^{93}Nb}$ & 1.4 $\times$ $\mathrm{10^{-14}}$ & 1 & 2.6 $\times$ $\mathrm{10^{-14}}$ & 6.6 $\times$ $\mathrm{10^{-14}}$ & & Determine its \textit{s}-process \\
& & & & & 9.5 & 2.0 $\times$ $\mathrm{10^{-14}}$ & 6.2 $\times$ $\mathrm{10^{-14}}$ & & contribution \\
& $\mathrm{^{94}Nb}$ & $\mathrm{\beta^-}$ & $\mathrm{^{94}Mo}$ & 1.1 $\times$ $\mathrm{10^{-12}}$ & 1 & 2.5 $\times$ $\mathrm{10^{-8}}$ & 9.3 $\times$ $\mathrm{10^{-7}}$ & & Compare to stardust \\
& & & & & 9.5 & 2.4 $\times$ $\mathrm{10^{-8}}$ & 9.3 $\times$ $\mathrm{10^{-7}}$ & & SiC grain data \\
\hline
$\mathrm{^{108}Cd}$ & & & & stable & & stable & stable & $\uparrow$ & \textit{p}-only \\
& $\mathrm{^{107}Pd}$ & $\mathrm{\beta^-}$ & $\mathrm{^{107}Ag}$ & 3.4 $\times$ $\mathrm{10^{-15}}$ & 1 & 1.2 $\times$ $\mathrm{10^{-14}}$ & 3.1 $\times$ $\mathrm{10^{-11}}$ & & Determine its \textit{s}-process \\
& & & & & 9.5 & 7.8 $\times$ $\mathrm{10^{-15}}$ & 2.9 $\times$ $\mathrm{10^{-11}}$ & & contribution \\
& $\mathrm{^{108}Ag}$ & $\mathrm{\beta^-}$ & $\mathrm{^{108}Cd}$ & 4.7 $\times$ $\mathrm{10^{-3}}$ & 1 & 4.9 $\times$ $\mathrm{10^{-3}}$ & 4.2 $\times$ $\mathrm{10^{-3}}$ & & \\
& & $\mathrm{\beta^+}$ & $\mathrm{^{108}Pd}$ & 1.4 $\times$ $\mathrm{10^{-4}}$ & 1 & 7.5 $\times$ $\mathrm{10^{-5}}$ & 1.3 $\times$ $\mathrm{10^{-5}}$ & & \\
& & & & & 9.5 & 1.4 $\times$ $\mathrm{10^{-4}}$ & 2.4 $\times$ $\mathrm{10^{-5}}$ & & \\
\hline
$\mathrm{^{129}I}$ & & $\mathrm{\beta^-}$ & $\mathrm{^{129}Xe}$ & 1.4 $\times$ $10^{-15}$ & 1 & 2.4 $\times$ $10^{-9}$ & 4.5 $\times$ $10^{-8}$ & $\downarrow$ & \textit{r}-only \\
& & & & & 9.5 & 1.9 $\times$ $10^{-9}$ & 4.1 $\times$ $10^{-8}$ & & Short-lived \\
& $\mathrm{^{128}I}$ & $\mathrm{\beta^-}$ & $\mathrm{^{128}Xe}$ & 4.3 $\times$ $\mathrm{10^{-4}}$ & 1 & 4.2 $\times$ $\mathrm{10^{-4}}$ & 2.8 $\times$ $\mathrm{10^{-4}}$ & & radionuclide \\
& & & & & 9.5 & 4.2 $\times$ $\mathrm{10^{-4}}$ & 2.8 $\times$ $\mathrm{10^{-4}}$ & & \\
\hline
$\mathrm{^{152}Gd}$ & & $\alpha$ & $\mathrm{^{148}Sm}$ & stable & & stable & stable & $\uparrow$ & \textit{p}-only \\
& $\mathrm{^{151}Sm}$ & $\mathrm{\beta^-}$ & $\mathrm{^{151}Eu}$ & 2.4 $\times$ $\mathrm{10^{-10}}$ & 1 & 5.0 $\times$ $\mathrm{10^{-10}}$ & 9.5 $\times$ $\mathrm{10^{-9}}$ & & Determine its \textit{s}-process \\
& & & & & 9.5 & 3.4 $\times$ $\mathrm{10^{-10}}$ & 8.0 $\times$ $\mathrm{10^{-9}}$ & & contribution \\
& $\mathrm{^{152}Eu}$ & $\mathrm{\epsilon}$ & $\mathrm{^{152}Sm}$ & 1.2 $\times$ $\mathrm{10^{-9}}$ & 1 & 6.9 $\times$ $\mathrm{10^{-9}}$ & 1.8 $\times$ $\mathrm{10^{-7}}$ & & \\
& & & & & 9.5 & 7.8 $\times$ $\mathrm{10^{-9}}$ & 1.1 $\times$ $\mathrm{10^{-6}}$ & & \\
& & $\mathrm{\beta^-}$ & $\mathrm{^{152}Gd}$ & 4.5 $\times$ $\mathrm{10^{-10}}$ & 1 & 3.2 $\times$ $\mathrm{10^{-8}}$ & 4.3 $\times$ $\mathrm{10^{-5}}$ & & \\
& & & & & 9.5 & 3.2 $\times$ $\mathrm{10^{-8}}$ & 4.2 $\times$ $\mathrm{10^{-5}}$ & & \\
\hline
$\mathrm{^{176}Hf}$ & & & & stable & & stable & stable & $\uparrow$ & Both are \textit{s}-only \\
& $\mathrm{^{176}Lu}$ & $\mathrm{\beta^-}$ & $\mathrm{^{176}Hf}$ & 5.8 $\times$ $\mathrm{10^{-19}}$ & 1 & 1.3 $\times$ $\mathrm{10^{-12}}$ & 1.3 $\times$ $\mathrm{10^{-7}}$ & & \\
& & & & & 9.5 & 1.3 $\times$ $\mathrm{10^{-12}}$ & 1.4 $\times$ $\mathrm{10^{-7}}$ & & \\
\hline
$\mathrm{^{182}Hf}$ & & $\mathrm{\beta^-}$ & $\mathrm{^{182}Ta}$ & 2.5 $\times$ $10^{-15}$ & 1 & 7.4 $\times$ $10^{-14}$ & 1.2 $\times$ $10^{-9}$ & $\downarrow$ & Short-lived \\
& & & & & 9.5 & 6.8 $\times$ $10^{-14}$ & 8.9 $\times$ $10^{-10}$ & & radionuclide \\
\hline
$\mathrm{^{205}Pb}$ & & $\mathrm{\epsilon}$& $\mathrm{^{205}Tl}$ & 1.3 $\times$ $10^{-15}$ & 1 & 1.7 $\times$ $10^{-9}$ & 3.1 $\times$ $10^{-10}$ & $\downarrow$\tablefootmark{a} & Single \textit{s}-only \\
& & & & & 9.5 & 6.4 $\times$ $10^{-9}$ & 1.9 $\times$ $10^{-9}$ & & short-lived \\
& $\mathrm{^{204}Tl}$ & $\mathrm{\beta^-}$ & $\mathrm{^{204}Pb}$ & 5.6 $\times$ $\mathrm{10^{-9}}$ & 1 & 6.3 $\times$ $\mathrm{10^{-9}}$ & 1.3 $\times$ $\mathrm{10^{-6}}$ & & radionuclide \\
& & & & & 9.5 & 6.3 $\times$ $\mathrm{10^{-9}}$ & 1.2 $\times$ $\mathrm{10^{-6}}$ & & \\
& $\mathrm{^{205}Tl}$ & $\mathrm{\beta_b^-}$ & $\mathrm{^{204}Pb}$ & stable & 1 & 1.1 $\times$ $\mathrm{10^{-12}}$ & 6.1 $\times$ $\mathrm{10^{-8}}$ & & \\
& & & & & 9.5 & 1.9 $\times$ $\mathrm{10^{-17}}$ & 2.0 $\times$ $\mathrm{10^{-8}}$ & & \\
\hline
\end{tabular}
\tablefoot{Columns indicate the corresponding isotope (Col. 1), the branching isotopes (Col. 2) that affect its abundance, their decay modes, daughter nuclei, and decay rates (Col. 3-8). $\lambda_{\mathrm{Set\,0}}$ indicates the decay rate used in the Set 0 models from JINA reaclib, while $\lambda^{\mathrm{90\,MK}}_{\mathrm{Set\,1}}$ and $\lambda^{\mathrm{300\,MK}}_{\mathrm{Set\,1}}$ indicate the value of the \textit{T}-dependent rate at 90 and 300 million K, respectively. The decay rate values are given for two different electron number densities, if available in the NETGEN database. The direction in which the abundance of each isotope has changed in the Set 1 models compared to the Set 0 models and the relevance of each case are shown in Cols. 9 and 10, respectively. \\
\tablefoottext{a}{Except for the \textit{M} = 4 $\mathrm{M_{\odot}}$ \textit{Z} = 0.007 model.}}
\end{table*}
\setlength{\tabcolsep}{1em}
\begin{table*}
\caption{$Y^\mathrm{i}_{\mathrm{Set\,2,final}}$/$Y^\mathrm{i}_{\mathrm{Set\,1,final}}$ for the 34 isotopes that vary the most between the two sets.}
\label{table:A.2.}
\centering
\begin{tabular}{cccccccc}
\hline
Isotope & \multicolumn{7}{c}{Stellar model} \\
& \multicolumn{2}{c}{\textit{Z} = 0.007} & \multicolumn{3}{c}{\textit{Z} = 0.014} & \multicolumn{2}{c}{\textit{Z} = 0.03} \\\cline{2-8}
& 3\,$\mathrm{M_{\odot}}$ & 4\,$\mathrm{M_{\odot}}$ & 2\,$\mathrm{M_{\odot}}$ & 3\,$\mathrm{M_{\odot}}$ & 4\,$\mathrm{M_{\odot}}$ & 3\,$\mathrm{M_{\odot}}$ & 4\,$\mathrm{M_{\odot}}$ \\
\hline
$\mathrm{^{64}Ni}$	&	\textbf{0.81}	&	1.00	&	\textbf{0.89}	&	\textbf{0.80}	&	0.94	&	\textbf{0.82}	&	\textbf{0.87}	\\
$\mathrm{^{65}Cu}$	&	\textbf{1.13}	&	\textbf{1.18}	&	1.08	&	\textbf{1.15}	&	\textbf{1.11}	&	\textbf{1.19}	&	\textbf{1.12}	\\
$\mathrm{^{70}Zn}$	&	1.06	&	\textbf{1.40}	&	1.02	&	1.06	&	1.07	&	1.05	&	1.03	\\
$\mathrm{^{74}Ge}$	&	\textbf{1.36}	&	\textbf{1.26}	&	\textbf{1.23}	&	\textbf{1.37}	&	\textbf{1.29}	&	\textbf{1.38}	&	\textbf{1.33}	\\
$\mathrm{^{75}As}$	&	\textbf{1.45}	&	\textbf{1.31}	&	\textbf{1.25}	&	\textbf{1.46}	&	\textbf{1.34}	&	\textbf{1.45}	&	\textbf{1.38}	\\
$\mathrm{^{78}Se}$	&	\textbf{1.12}	&	1.00	&	\textbf{1.15}	&	\textbf{1.17}	&	1.08	&	\textbf{1.19}	&	\textbf{1.14}	\\
$\mathit{^{79}Se}$	&	\textit{1.05}	&	\textit{0.91}	&	\textit{\textbf{1.14}}	&	\textit{\textbf{1.11}}	&	\textit{1.01}	&	\textit{\textbf{1.15}}	&	\textit{\textbf{1.10}}	\\
$\mathrm{^{81}Br}$	&	\textbf{1.41}	&	\textbf{1.29}	&	\textbf{1.26}	&	\textbf{1.40}	&	\textbf{1.37}	&	\textbf{1.37}	&	\textbf{1.36}	\\
$\mathrm{^{85}Rb}$	&	1.06	&	1.00	&	1.09	&	\textbf{1.10}	&	1.05	&	1.09	&	1.06	\\
$\mathrm{^{89}Y}$	&	\textbf{1.13}	&	1.05	&	\textbf{1.15}	&	\textbf{1.13}	&	\textbf{1.10}	&	1.06	&	1.03	\\
$\mathit{^{94}Nb}$	&	\textit{\textbf{0.74}}	&	\textit{1.01}	&	\textit{\textbf{0.79}}	&	\textit{0.99}	&	\textit{1.03}	&	\textit{0.94}	&	\textit{\textbf{0.89}}	\\
$\mathrm{^{110}Cd}$	&	0.96	&	0.96	&	0.98	&	0.94	&	0.93	&	0.91	&	\textbf{0.89}	\\
$\mathrm{^{121}Sb}$	&	\textbf{1.16}	&	1.08	&	\textbf{1.13}	&	\textbf{1.10}	&	\textbf{1.10}	&	1.05	&	1.04	\\
$\mathrm{^{122}Te}$	&	0.96	&	\textbf{0.90}	&	0.93	&	0.91	&	0.93	&	\textbf{0.88}	&	\textbf{0.90}	\\
$\mathit{^{129}I}$	&	\textit{1.09}	&	\textit{1.01}	&	\textit{\textbf{1.11}}	&	\textit{1.07}	&	\textit{1.05}	&	\textit{1.07}	&	\textit{1.01}	\\
$\mathrm{^{132}Xe}$	&	\textbf{1.17}	&	\textbf{1.16}	&	\textbf{1.14}	&	\textbf{1.11}	&	\textbf{1.13}	&	1.07	&	1.05	\\
$\mathrm{^{134}Ba}$	&	0.96	&	0.98	&	0.91	&	\textbf{0.90}	&	0.97	&	\textbf{0.87}	&	0.92	\\
$\mathrm{^{136}Ba}$	&	0.90	&	0.95	&	\textbf{0.90}	&	\textbf{0.87}	&	0.93	&	\textbf{0.87}	&	\textbf{0.88}	\\
$\mathrm{^{137}Ba}$	&	\textbf{0.81}	&	1.02	&	\textbf{0.74}	&	\textbf{0.75}	&	0.97	&	\textbf{0.75}	&	\textbf{0.84}	\\
$\mathrm{^{140}Ce}$	&	\textbf{1.12}	&	\textbf{1.10}	&	\textbf{1.11}	&	\textbf{1.12}	&	1.08	&	1.08	&	1.03	\\
$\mathrm{^{142}Ce}$	&	\textbf{1.14}	&	\textbf{1.17}	&	1.01	&	1.09	&	\textbf{1.13}	&	1.01	&	1.02	\\
$\mathrm{^{148}Sm}$	&	0.95	&	1.00	&	\textbf{0.89}	&	0.91	&	0.97	&	\textbf{0.89}	&	0.92	\\
$\mathrm{^{152}Gd}$	&	0.91	&	0.94	&	\textbf{0.86}	&	\textbf{0.88}	&	0.93	&	\textbf{0.86}	&	0.92	\\
$\mathrm{^{160}Dy}$	&	0.93	&	0.98	&	\textbf{0.89}	&	0.91	&	0.95	&	\textbf{0.89}	&	0.92	\\
$\mathrm{^{170}Er}$	&	\textbf{1.26}	&	\textbf{1.47}	&	1.04	&	\textbf{1.15}	&	\textbf{1.33}	&	1.02	&	1.05	\\
$\mathrm{^{172}Yb}$	&	\textbf{0.90}	&	0.97	&	\textbf{0.88}	&	\textbf{0.90}	&	0.95	&	\textbf{0.89}	&	0.93	\\
$\mathrm{^{174}Yb}$	&	\textbf{0.90}	&	0.98	&	\textbf{0.90}	&	\textbf{0.90}	&	0.95	&	\textbf{0.90}	&	0.92	\\
$\mathrm{^{179}Hf}$	&	\textbf{0.90}	&	0.97	&	0.92	&	0.91	&	0.95	&	0.92	&	0.94	\\
$\mathrm{^{181}Ta}$	&	\textbf{0.90}	&	0.96	&	0.92	&	0.92	&	0.94	&	0.92	&	0.93	\\
$\mathrm{^{186}W}$	&	\textbf{1.27}	&	\textbf{1.44}	&	\textbf{1.12}	&	\textbf{1.19}	&	\textbf{1.32}	&	1.07	&	1.07	\\
$\mathit{^{187}Re}$	&	\textit{0.99}	&	\textit{\textbf{1.11}}	&	\textit{0.97}	&	\textit{0.97}	&	\textit{1.03}	&	\textit{0.99}	&	\textit{0.99}	\\
$\mathrm{^{192}Os}$	&	\textbf{1.98}	&	\textbf{2.30}	&	\textbf{1.10}	&	\textbf{1.30}	&	\textbf{1.37}	&	1.04	&	1.03	\\
$\mathrm{^{193}Ir}$	&	1.03	&	\textbf{1.11}	&	1.00	&	1.00	&	1.03	&	1.00	&	1.00	\\
$\mathrm{^{197}Au}$	&	\textbf{0.90}	&	1.00	&	0.96	&	0.94	&	0.97	&	0.97	&	0.99	\\
\hline
\end{tabular}
\tablefoot{$Y^\mathrm{i}_{\mathrm{Set\,2,final}}$/$Y^\mathrm{i}_{\mathrm{Set\,1,final}}$ indicates the ratio of the final surface normalised number abundance of isotope \textit{i} between Set 2 and Set 1. We list only those isotopes that abundance changed at least 10\% (in bold) in at least one Set 2 model compared to the same Set 1 model. The unstable isotopes are in italics.}
\end{table*}
\end{appendix}
\end{document}